\documentclass[aps,floatfix,onecolumn,notitlepage,superscriptaddress,preprintnumbers,nofootinbib,11pt]{revtex4-2}
\usepackage{natbib,setspace}
\usepackage{graphicx,tikz-feynman}
\usepackage{amssymb}
\usepackage{amsmath,amssymb,stix}
\usepackage{color}
\usepackage{caption}
\usepackage{subcaption}
\usepackage{slashed}

\usepackage{fancyhdr}
\definecolor{darkblue}{rgb}{0,0,0.5}
\usepackage[colorlinks,linkcolor=darkblue,citecolor=darkblue,urlcolor=darkblue]{hyperref}

\usepackage[capitalize]{cleveref}
\allowdisplaybreaks

\newcommand\beq{\begin{equation}}
\newcommand\eeq{\end{equation}}

\begin{document}

%%%%%%%%%%%%%%%%%%%%%%%%%%%%%%%%%%%%%%%%%%%%%%%%%%%%%%%%%%%%%%%%%%

\title{Search for $Z^\prime$ Radiating from the Dark Matter at the LHC}

%\title{The 146 GeV $H\to e\mu$ excess}
%%%%%%%%%%%%%%%%%%%%%%%%%%%%%%%%%%%%%%%%%%%%%%%%%%%%%%%%%%%%%%%%%%

\author{Akmal Ferdiyan}
\email{aferdiyan@unsoed.ac.id}
\affiliation{Theoretical High Energy Physics Research Division,
Faculty of Mathematics and Natural Sciences, Institut Teknologi Bandung,
Jl. Ganesha no.10 Bandung, 40132, Indonesia }
\author{Reinard Primulando}
\email{rprimulando@unpar.ac.id}
\affiliation{Center for Theoretical Physics, Department of Physics, Parahyangan Catholic University, Jalan Ciumbuleuit 94, Bandung 40141, Indonesia}
\author{Qidir Maulana Binu Soesanto}
\email{qidirbinu@fisika.fsm.undip.ac.id}
\affiliation{Department of Physics, Faculty of Sciences and Mathematics, Universitas Diponegoro,
Jl. Prof. Dr. Sudarto No.13, Semarang, Indonesia, 50275}
\author{Bayu Dirgantara}
\email{bayuquarkquantum@yahoo.com}
\affiliation{School of Physics and Center of Excellence in High Energy Physics and Astrophysics, Suranaree University of Technology, Nakhon Ratchasima
30000, Thailand}
\author{Bobby Eka Gunara}
\email{bobby@itb.ac.id}
\affiliation{Theoretical High Energy Physics Research Division,
Faculty of Mathematics and Natural Sciences, Institut Teknologi Bandung,
Jl. Ganesha no.10 Bandung, 40132, Indonesia }

%%%%%%%%%%%%%%%%%%%%%%%%%%%%%%%%%%%%%%%%%%%%%%%%%%%%%%%%%%%%%%%%%%
\begin{abstract}
We discuss a collider probe of a dark sector model in which the dark matter is charged under a new, hidden $U(1)$ gauge group. In particular, we look for the so-called Darkstrahlung process, in which the final states dark matter radiates a new $Z^\prime$ gauge boson and it manifests as dilepton resonances. This work emphasizes the potential of dilepton final states with missing transverse energy in probing the darkstrahlung process. We recast the ATLAS Run 2 search for dilepton resonances in association with missing energy in addition to the ATLAS and CMS searches for sleptons. We find that the recasted searches put strong constraints on the coupling between the dark matter and the $Z^\prime$. Moreover, we evaluate refined search strategies for $Z^\prime$ production and propose an analysis employing constraints on the lepton invariant mass and higher missing energy cut related to the darkstrahlung process. Simulation outcomes indicate substantial enhancements, particularly by a factor of 6 in regions featuring lower $Z^\prime$ masses. Finally, we also discuss the case when the $Z^\prime$ has a long lifetime, resulting in displaced decay of the boson.
\end{abstract}
%%%%%%%%%%%%%%%%%%%%%%%%%%%%%%%%%%%%%%%%%%%%%%%%%%%%%%%%%%%%%%%%%%

\maketitle
\newpage

\flushbottom
%%%%%%%%%%%%%%%%%%%%%%%%%%%%%%%%%%%%%%%%%
\section{Introduction}
%%%%%%%%%%%%%%%%%%%%%%%%%%%%%%%%%%%%%%%%

Several cosmological observations strongly suggest the existence of non-baryonic dark matter (DM), which appears to dominate over baryonic matter in the universe. However, our understanding of the properties of dark matter particles remains limited. The Large Hadron Collider (LHC) plays an integral role in illuminating the properties of dark matter. The LHC is aptly suited for scrutinizing the coupling of dark matter with quarks. Furthermore, if such coupling exists, it may manifest detectable signals in direct detection experiments. Hence a comparison analysis can be done between collider and direct detection experiments. To capture the kinematics of the collider, a simplified model is commonly used in which the dark matter coupling is mediated with a heavy mediator~\cite{Albert:2022xla,Buchmueller:2014yoa,DEramo:2016gos,Duerr:2016tmh,Cohen:2021gdw}. One frequently utilized model involves a heavy neutral vector mediator, denoted as $Z_A$, which interacts with both quarks and dark matter, leading to dark matter production via the $s$-channel exchange of the mediator.

Since the dark matter itself is invisible at the LHC, it has to recoil against some other visible objects to be detectable. The visible objects can come from the initial quarks or gluons radiating jets~\cite{Goodman:2010ku,Fox:2011pm,Lin:2013sca,Haisch:2015ioa}, photons~\cite{Fox:2011pm}, $W$-bosons~\cite{Bell:2015rdw,Bai:2012xg}, $Z$-bosons~\cite{Bell:2012rg,Carpenter:2012rg} or Higgses~\cite{Petrov:2013nia,Berlin:2014cfa}, resulting in signatures commonly referred to as mono-$X$. Amongst these possible signatures, monojet events impose the strongest constrain on the simplified model. For instance, ATLAS collaboration \cite{ATLAS:2021kxv}, utilizing all the LHC run 2 data, established exclusion limits on the mediator $Z_A$ up to the order of TeV for dark matter masses up to 500 GeV. While monojet searches provide robust constraints, they also face substantial backgrounds, primarily from the irreducible $Z(\rightarrow\nu\bar\nu)+$jets backgrounds. Even within signal regions with missing transverse energy exceeding 200 GeV, the search conducted by \cite{ATLAS:2021kxv} allows for the production of around $10,000$ dark matter events without exclusion.

 It is conceivable that the dark sector is governed by dark gauge groups, wherein dark matter may radiate dark gauge bosons in the final states. If the dark gauge bosons decay back to the invisible dark sector, the collider signature would resemble previous mono-$X$ searches. However, in numerous models, dark gauge bosons exhibit mixing with Standard Model (SM) gauge bosons, often through kinetic mixing mechanisms~\cite{Strassler:2006im,Han:2007ae,Holdom:1985ag}. In scenarios where dark gauge bosons cannot kinematically decay into other dark sector particles, they decay into SM particles regardless of the strength of the mixing. Consequently, a distinctive signature emerges: a pair of SM particles accompanied by missing energy, with the invariant mass of the SM particles reconstructing the mass of the dark gauge boson. This phenomenon, termed the ``Darkstrahlung'' process, was discussed in \cite{Gupta:2015lfa}. Given the abundant production of dark matter at the LHC without exclusion by monojet searches, darkstrahlung presents a promising avenue for uncovering the properties of dark matter. Several other proposals on radiating mediators or dark gauge bosons from the invisible dark sector at the LHC can be found in Refs.~\cite{Bai:2015nfa,Tsai:2015ugz,Lindner:2016lpp,Buschmann:2016hkc,Kim:2016fdv,Nam:2021bsf}. There are also discussion on the effect of radiating dark gauge boson in the DM direct searches~\cite{Kim:2019had} and indirect searches~\cite{Bell:2017irk}.

In this paper we will update the previous works of~\cite{Gupta:2015lfa} by incorporating the latest LHC run-2 data. A recent work by ATLAS~\cite{ATLAS:2023tmv} focuses on a specific darkstrahlung signature: a resonant lepton pair accompanied by missing transverse energy\footnote{Earlier Ref.~\cite{Elgammal:2021rne} utilized the CMS Open Data with 11.6 fb$^{-1}$ at $\sqrt{s} = 8$ TeV to search for the resonant muon pair associated with missing transverse energy.}. However, this analysis adopts a different benchmark model based on Ref.~\cite{Bai:2015nfa}. We will recast the search to incorporate the darkstrahlung process. Additionally, we will recast the ATLAS~\cite{ATLAS:2019lff} and CMS~\cite{CMS:2020bfa} searches of the slepton. The ATLAS and CMS slepton searches are looking for two non-resonant leptons with a significant presence of missing energy. Although the slepton search channel is not optimized for darkstrahlung, it can still yield stringent bounds. Indeed, our findings indicate that slepton searches may surpass the sensitivity of the ATLAS search \cite{ATLAS:2023tmv}, which targets a resonant lepton pair. We will also explore methods to optimize these searches for darkstrahlung. Furthermore, we will discuss constraints in scenarios where the dark gauge boson exhibits an extended lifetime, due to the smallness of the mixing.

Our paper is structured as follows: In Section \ref{sec:model}, we will explain the model used in this work. Section \ref{sec:prompt} discusses the case of the bounds on promptly decaying $Z'$. The case of long lived $Z'$ is discussed in Section \ref{sec:LLP}. Finally we conclude in Section \ref{sec:conc}.

%%%%%%%%%%%%%%%%%%%%%%%%%%%%%%%%%%%%%%%%%
\section{The Model}
\label{sec:model}

In this work, we will investigate a model featuring fermionic dark matter $\chi$, with corresponding Abelian $U(1)$ gauge group, $U(1)_D$. The interaction of dark matter is described by a coupling strength, $g_D$, under the dark $U(1)_D$ gauge group, represented by the Lagrangian term:
\begin{equation}
    \mathcal{L}_{\text{Dark}} = g_D X_{\mu} \Bar{\chi} \gamma^{\mu} \chi,
\end{equation}
where $X_{\mu}$ denotes the dark gauge boson field. 

Additionally, the dark $U(1)_D$ kinetically mixes with the $U(1)_Y$ gauge field of the SM. The kinetic mixing term is expressed as
\begin{equation}
    \mathcal{L}_{\text{kin.mixing}} = - \frac{1}{4} B_{\mu \nu}B^{\mu \nu}+\frac{\epsilon}{2 \cos\theta_W}B_{\mu \nu}X^{ \mu \nu} - \frac{1}{4} X_{\mu \nu}X^{\mu \nu} 
\end{equation}
where $B_{\mu\nu}$ and $X_{\mu\nu}$ are the field strength tensors of the $U(1)_Y$ and $U(1)_D$, respectively. Here, the strength of the kinetic mixing is represented by $\epsilon$ normalized by $\cos\theta_{W}$, with $\theta_W$ being the weak mixing angle. For our purpose, we introduce the following notations:
\begin{align}
    \kappa = & \frac{\epsilon}{\cos\theta_W}, \\
    \eta = & \frac{\kappa}{\sqrt{1-\kappa^2}},
\end{align}
which is also used in other sources, such as in \cite{San:2021zpeps}. There exist bounds on the values of $\epsilon$ coming from the direct productions of the $Z'$ at colliders~\cite{CMS:2020epss} or from electroweak precision variables~\cite{Curtin:2014cca}. In general, the upper limits on $\epsilon$ are in the order of $\mathcal{O}(10^{-3})$.

The mass of the dark gauge boson can be generated through mechanisms like spontaneous breaking by a dark Higgs or the Stueckelberg mechanism \cite{Bell:2016uhg}. The gauge boson masses are expressed as
\begin{equation}
    \mathcal{L}_{\text{mass}} = -\frac{v^{2}}{4}\left(g_{2}W_{\mu}^{3}-g_{1}B_{\mu}\right) \left(g_{2}W^{\mu}_{3}-g_{1}B^{\mu}\right)-\frac{m_{X}^{2}}{2}X_{\mu}X^{\mu},
\end{equation}
where $W_\mu^3$ is the third component of the $SU(2)_L$ gauge boson and $v$, $g_1$, $g_2$, and $m_{X}$ are the VEV of scalar Higgs boson, the $SU(2)_L$ gauge coupling, the $U(1)_Y$ gauge coupling, and the Lagrangian term of the mass of gauge boson $X_\mu$, respectively. Canonically normalized kinetic terms can be obtained by performing orthogonal rotation, propagating to the mass and interaction terms. Subsequent rotation yields the mass eigenstate, with two resulting massive gauge bosons: the SM gauge boson denoted by $Z$ and the dark gauge boson denoted by $Z'$. %Note that the last r leaves the canonical kinetic term invariant. %These gauge bosons are not expressed in their mass eigenstates, %gauge eigenstates, 
%further diagonalization is needed. After diagonalization, we will have the mass of the dark gauge boson, $m_{Z'}$ and . 

\begin{figure}
    \centering
    \includegraphics{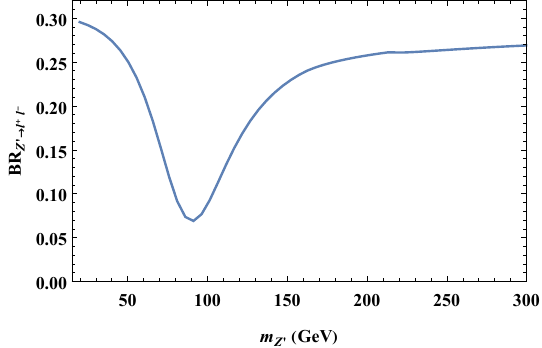}
    \caption{Branching fraction of $Z'$ to two leptons ($e^+e^-$ and $\mu^+\mu^-$) as a function of mass.}
    \label{fig:br_zprime}
\end{figure}

Due to the rotations on the $X_\mu$, $B_\mu$ and $W^3_\mu$, the dark $Z'$ is allowed to interact with both SM particles and dark particles. However in the scenario in which ${m_{Z'}} < 2 m_{\chi}$, $Z'$ will decay exclusively into SM particles, regardless of the value of the mixing parameter, $\epsilon$. The decay width of $Z'$ to fermion pairs is given by  
\begin{equation}
    \Gamma_{Z'\to f^{+} f^{-}} = \frac{m_{Z'}}{12 \pi} N_c \ \sqrt{1 - \frac{4 m_f^2}{m_{^{Z'}}^2}} \left[ \left( c^{ 2}_V + c^{
  2}_A \left) \left( 1 - \frac{m_f^2}{m_{^{Z'}}^2} \right) + 3 (c^{ 2}_V
  - c^{ 2}_A) \frac{m_f^2}{m_{^{Z'}}^2} \right] \right. \right.,
    \label{decaywidth}
\end{equation}
where $N_c$ is the color factor and $m_f$ is the mass of the fermions in the final state. The coefficients $c_V$ and $c_A$ are given by 
\begin{align}
   c_A &= -\frac{g_2}{\cos\theta_W}\frac{T_{f}^3}{2} (\sin\alpha + \eta \sin\theta_W \cos\alpha), \\ 
    c_V &= -\frac{g_2}{\cos\theta_W} \left(\frac{T_{f}^3}{2} - Q_f \sin\theta_W^2\right)(\sin\alpha + \eta \sin\theta_W \cos \alpha)  +\eta e Q_f \cos\theta_W \cos\alpha,
\end{align}
where $T^3_f$ is the weak isospin of $f$, $Q_f$ is the charge of $f$ and $e \equiv g_2 \sin\theta_W$. The rotation angle $\alpha$ is given by 
\begin{equation}
    \tan 2\alpha = \frac{-2 \eta \sin\theta_W}{1-{c^2_X} - \eta^2 ({c^2_X}+{\sin^2\theta_W})}, 
\end{equation}
with
\begin{equation}
    c_X = \sqrt{\frac{{m^2_{Z'}} ({M^2_Z} - {m^2_{Z'}} + {M^2_Z} ~\eta^2 ~{\sin^2\theta_W}  )}{{M^2_Z} ({M^2_Z}-{m^2_{Z'}})(1+\eta^2)}}.
\end{equation}
Here $m_{Z'}$ is the physical mass of $Z'$ boson, while $M_Z$ is the unrotated mass term for $Z$ boson defined as \mbox{${M_Z} = \frac{1}{2}v\sqrt{g_1^2+g_2^2}$}.
The branching fraction of $Z'$ to leptons pair ($e^+e^-$ and $\mu^+\mu^-$) is depicted in Fig.~\ref{fig:br_zprime}. Note that the branching fraction is independent of $\epsilon$, as long as $m_{Z'} < 2 m_\chi$.

%This will lead to a detectable final state resonance in conjunction with considerable missing energy originating from the stable particles $\chi$.

Finally, we will use a simplified model for dark matter productions at the LHC similar to the ones used in the LHC monojet analysis \cite{ATLAS:2021kxv}. The dark matter couples through quarks via a heavy mediator $Z_A$,
\begin{equation} \label{eq:monojet}
    \mathcal{L}_{q-\chi} = g_Q (\Bar{u} \gamma^{\mu} \gamma^{5} u + \Bar{d}\gamma^{\mu} \gamma^{5} d) {Z_{A}}^{\mu} + g_{\chi} \Bar{\chi} \gamma^{\mu} \gamma^{5} \chi {Z_{A}}^{\mu},
\end{equation}
with $g_Q$ and $g_\chi$ are free coupling parameters. The Lagrangian $\mathcal{L}_{q-\chi}$ produces monojet signatures at the LHC and the search~\cite{ATLAS:2021kxv} produces $\mathcal{O}(\text{TeV})$  bounds on the $m_{Z_A}$ for $g_Q = 0.25$ and $g_\chi = 1$.

\section{Prompt $Z'$ Search}
\label{sec:prompt}

\subsection{Search Strategy}

%The dark matter, $\chi$, can be produced by the s-channel mediator. The production is probed by the monojet searches, such as \cite{ATLAS:2021kxv}. 
The  $Z'$ can be observed at various resonance channels, such as dijet, ditau or dilepton resonances. Among these, the dilepton resonance channel \cite{ATLAS:2019erb, CMS:2021ctt} stands out due to its significant branching fraction to leptons and relatively low background at the LHC. However, for $Z'$ masses on the order of $\mathcal{O}(100  \text{ GeV})$, the background is already at around $10,000$ events/GeV. This large value of the backgrounds demands exploring other search channels with lower backgrounds, especially when $Z'$ is produced with a small cross section. 

The production of $Z'$ can occur via different mechanisms including Drell-Yan production channel through its mixing with $U(1)_Y$ gauge field. However, for the small values of mixing parameter $\epsilon$, the cross section becomes negligible at the LHC scale. Nonetheless, there is another viable production mechanism involving the emission of $Z'$ by the final state $\chi$. 

\begin{figure}[h!]
	\begin{center}
		\begin{tikzpicture}
			\begin{feynman}
				\vertex (a);
				\vertex [above left=of a] (b);
				\vertex [below left=of a] (c);
				\vertex [above left=of b] (bb);
				\vertex [below left=of c] (cc);
				\vertex [right=of a] (d);
				\vertex [above right=of d] (e);
				\vertex [above right=of e] (f);
				\vertex [right=of e] (g);
				\vertex [above right=of g] (h);
				\vertex [below right=of g] (i);
				\vertex [below right=of d] (j);
				\vertex [below right=of j] (k);
				
				\diagram{
					(bb) -- [fermion, edge label=$q$](a) -- [fermion, edge label=$\bar{q}$] (cc);
					(a) -- [boson, edge label=$Z_A$] (d);
					(d) -- [fermion, edge label=$\chi$] (e) -- [fermion, edge label=$\chi$] (f);
					(d) -- (j);
                    (k) -- [fermion, edge label=$\bar\chi$] (j);
					(e) -- [boson, edge label=$Z'$] (g);
					(h) -- [fermion, edge label=$\ell^+$] (g);
					(g) -- [fermion, edge label=$\ell^-$] (i);
				};
			\end{feynman}
		\end{tikzpicture}
	\end{center}
	\caption{The darkstrahlung process.}
 \label{feynmandiag}
\end{figure}
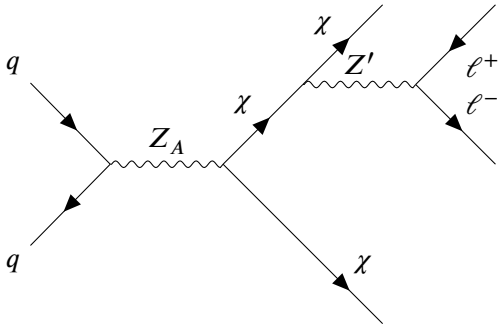

\begin{figure}[h]
\includegraphics[width=0.7\textwidth]{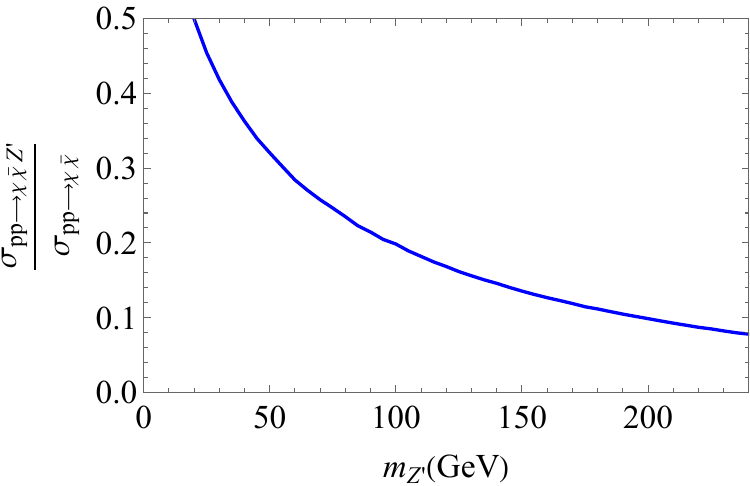}
\caption{The ratio of $\sigma_{pp \rightarrow \chi \bar\chi Z'}$ and $\sigma_{pp \rightarrow \chi \bar\chi }$ as a function of $m_{Z'}$ for $m_{Z_A} = 2$ TeV, $m_\chi = 120$ GeV and $g_D = 1$. The analytical formula for the ratio of related production channels $q \bar q \rightarrow \chi\chi Z'$ and $q \bar q \rightarrow \chi\chi$ is discussed in Appendix~\ref{sec:app}.}
\label{fig:xsratio}
\end{figure}

While the monojet analysis \cite{ATLAS:2021kxv} provides significant bounds on the production of dark matter, $\chi$, it is not a zero background process. In fact, for the missing energy greater than 200 GeV, the search allows an upper limit of $\mathcal O(10^5)$ dark matter production events at the 95\% confidence level (C.L.), implying that a substantial number of dark matter events might go undetected. When the $Z'$ coupling to the dark matter being particularly large, a considerable portion of the produced dark matter can radiate $Z'$, termed darkstrahlung, as depicted in Fig. ~\ref{feynmandiag}. The ratio of darkstrahlung cross section to the dark matter pair production cross-section, shown in Fig. ~\ref{fig:xsratio}, suggests that darkstrahlung events can still be produced in significant amount at the LHC.  Consequently, our focus is directed towards a process wherein dark matter radiates $Z'$, succeeded by the subsequent decay of $Z'$ into lepton pairs\footnote{For relatively light $Z'$, the signature would be lepton jets coming from the $Z'$ decay as discussed in Ref.~\cite{Buschmann:2015awa}}.

The exploration of a dilepton resonance channel with a substantial missing energy component presents a unique, low-background opportunity for probing this dark sector. 
A recent ATLAS search on this signature~\cite{ATLAS:2023tmv} provides a platform for reinterpreting the search to constrain the darkstrahlung process. The search utilizes 140 fb$^{-1}$ of the LHC run 2 data with the corresponding $\sqrt{s} = 13$ TeV. The ATLAS search requires the events to have two oppositely charged light leptons with same flavor. The electrons must satisfy the requirements of $p_T>25$ GeV and $|\eta|<2.47$, but outside of the range  1.37 < $\lvert\eta\rvert$ < 1.52. The muons are required to have $p_T>25$ GeV and $|\eta|<2.5$. The jets must meet the criteria of $p_T>20$ GeV and $|\eta|<4.5$. Events are also constrained to have $E_T^\text{miss} >55$ GeV and $m_{ll} > 180$ GeV, with any $b$-tagged jet is rejected. ATLAS collaboration separates the signal region into three regions based on the $E_T^\text{miss}$ significance. Delphes detector simulator~\cite{deFavereau:2013fsa} cannot simulate the $E_T^\text{miss}$ significance parameter, hence in this analysis we combine the three signal regions into one region. We expect that resulting bounds would be weaker than having the signal regions separated.

Since the dilepton resonance search only covers $m_{ll} > 180$ GeV, we will also recast the slepton search that seeks for two non-resonant, opposite sign leptons with missing energy. We adopt the selection criteria from 13 TeV ATLAS search for slepton~\cite{ATLAS:2019lff} that is based on 139 fb$^{-1}$ of collision data at $\sqrt{s} = 13$ TeV. The signal region is defined by two same flavor and oppositely charged leptons. The electrons are required to have $p_T>25$ GeV and $|\eta|<2.47$ while the muons must satisfy $p_T > 25$ GeV and $|\eta|<2.7$. Only the jets with $p_T>20$ GeV and $|\eta|<2.4$ are considered. The dilepton invariant mass, $m_{ll}$, is constrained to be $>121.2$ GeV. Events must satisfy the requirements of $E_T^\text{miss}>110$ GeV. The events with any reconstructed $b$-tagged jets are removed and all events must have no more than one non $b$-tagged jet. The events is then classified by the number of the non $b$-tagged jets: $n_j=0$ and $n_j=1$ events.

For lower bounds on $m_{Z'}$, a recast of CMS search for dilepton final states and MET, based on \cite{CMS:2020bfa}, is also performed. This search uses data sample of proton-proton collisions at $\sqrt{s}= 13$ TeV and corresponding to an integrated luminosity of 137 fb$^{-1}$. For CMS search, the event selection requires two opposite sign (OS) leptons with $|\eta|<2.4$ and $p_T> 50 (20) $ GeV for the highest (next-to-highest) $p_T$. Jets must satisfy the requirements of $p_T>20$ GeV and $|\eta|<2.4$. Other criteria are $20 \text{ GeV} <m_{ll} < 65$ GeV or~ $m_{ll}>120$ GeV, and $\text{m}_{\text{T2}}(ll) >100$ GeV. Additionally, the value of $E_T^\text{miss}$ must be greater than $50$ GeV.  This search also rejects any events with $b$-tagged jets. The multiplicity of jets is used to classify the events into $n_j=0$ and $n_j>0$ categories. 

To compare existing bounds from the experiments with our model's signal, we utilized  specific combinations of the $m_\chi$ and $m_{Z_A}$ with the values equal to the the 95\% C.L. bounds obtained in the monojet analysis~\cite{ATLAS:2021kxv}. Subsequently, with $m_{Z'}$ set at a fixed value, the resulting constraints on the dark coupling constant ($g_D$) will be examined.

%We would like to compare the existing bounds from ATLAS and CMS for dilepton + missing ET final state with our signal from our model. For this purpose, we will assign fixed values for parameters in our model and look at the constraint for the value of the dark coupling constant, $g_D$. 

To obtain background and signal events, Monte Carlo simulations are conducted using MadGraph v5~\cite{Alwall:2014hca} for parton-level event generation. Standard Model UFO in MadGraph generates background samples, while signal events are simulated using FeynRules 2.0~\cite{Alloul:2013bka}. Parton showering and hadronization are performed using Pythia 8.3~\cite{Sjostrand:2014zea}, and Delphes v3~\cite{deFavereau:2013fsa} is utilized for simulating detector effects. Jets are reconstructed employing the anti-$k_T$ algorithm with a radius parameter $R=0.4$, by using the FastJet~\cite{Cacciari:2012fj} implementation integrated into Delphes v3. 

\begin{figure}
     \centering
     \begin{subfigure}[b]{0.48\textwidth}
         \centering
         \includegraphics[width=\textwidth]{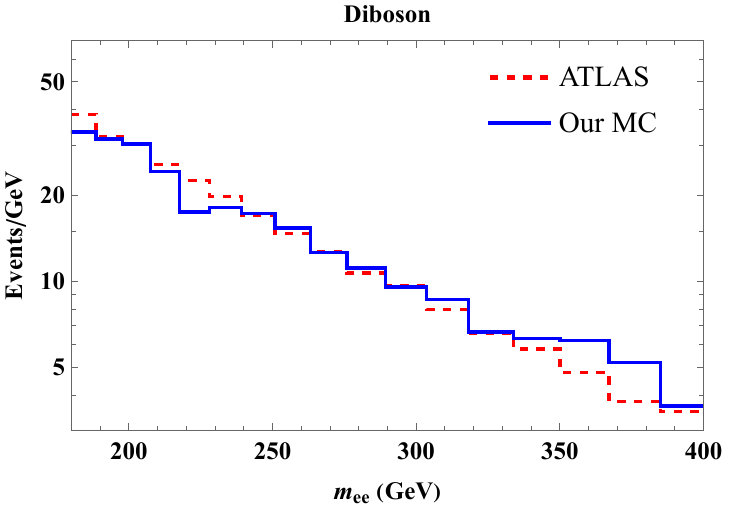}
         \caption{$ee$ channel}
         \label{fig:dibosonee}
     \end{subfigure}
     \hfill
     \begin{subfigure}[b]{0.48\textwidth}
         \centering
         \includegraphics[width=\textwidth]{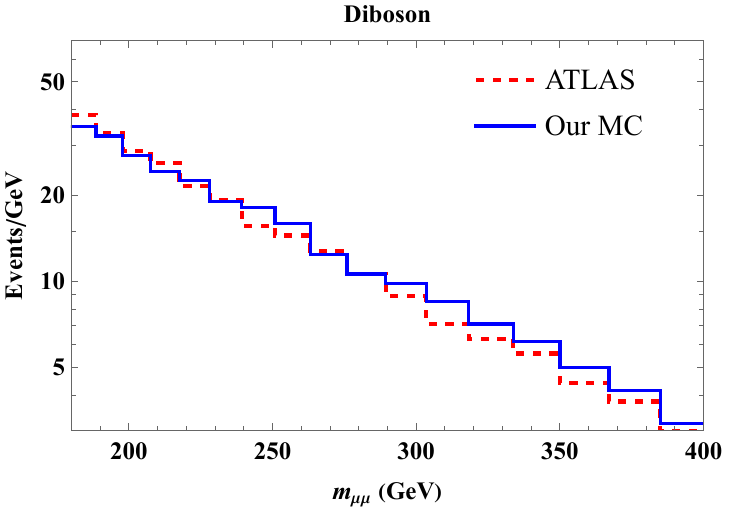}
         \caption{$\mu\mu$ channel}
         \label{fig:dibosonmumu}
     \end{subfigure}
\caption{Comparison between the ATLAS background prediction and our simulated diboson background from the dilepton resonance search~\cite{ATLAS:2023tmv}.}
\label{fig:comparisonWW}
\end{figure}

\begin{figure}
     \centering
     \begin{subfigure}[b]{0.48\textwidth}
         \centering
         \includegraphics[width=\textwidth]{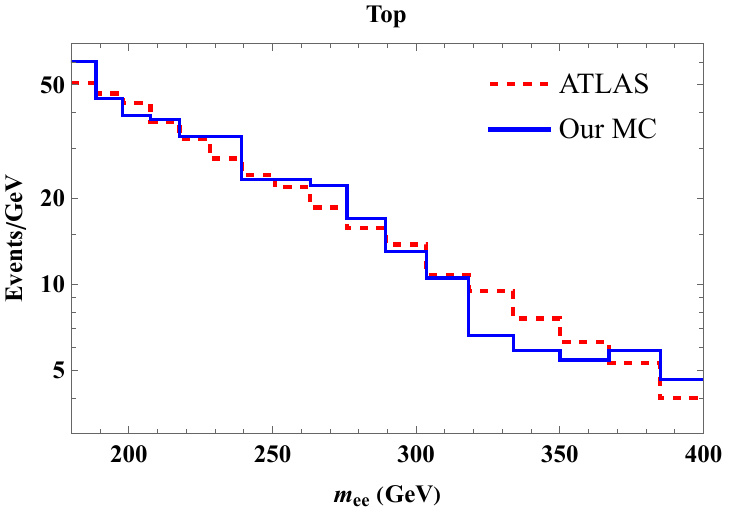}
         \caption{$ee$ channel}
         \label{fig:topee}
     \end{subfigure}
     \hfill
     \begin{subfigure}[b]{0.48\textwidth}
         \centering
         \includegraphics[width=\textwidth]{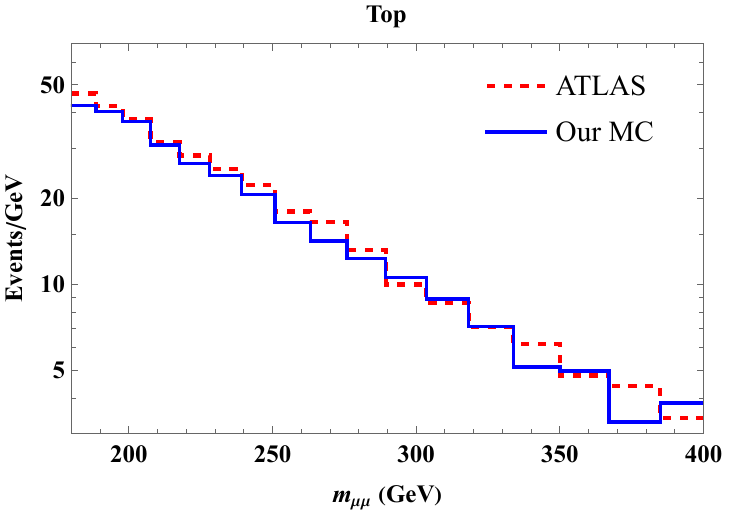}
         \caption{$\mu\mu$ channel}
         \label{fig:topmumu}
     \end{subfigure}
\caption{Comparison between the ATLAS background prediction and our simulated top background from the dilepton resonance search~\cite{ATLAS:2023tmv}.}
        \label{fig:comparisontt}
\end{figure}

The main background for the dilepton resonance search~\cite{ATLAS:2023tmv} comes from diboson ($WW$ and $ZZ$) and also $t\bar t$. We show the comparison between ATLAS estimates and our MC simulation for both backgrounds in Fig. \ref{fig:comparisonWW} and \ref{fig:comparisontt}. From the plots, we can see that the backgrounds are well reproduced. Note that  we have combined the number of events from the three signal regions in the plots. 

\begin{figure}
     \centering
     \begin{subfigure}[b]{0.48\textwidth}
         \centering
         \includegraphics[width=\textwidth]{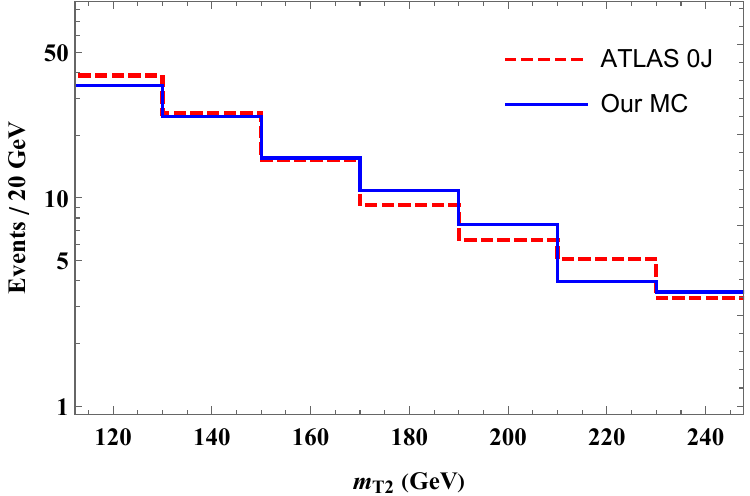}
         \caption{0J channel}
         \label{fig:slepton 0J MC}
     \end{subfigure}
     \hfill
     \begin{subfigure}[b]{0.48\textwidth}
         \centering
         \includegraphics[width=\textwidth]{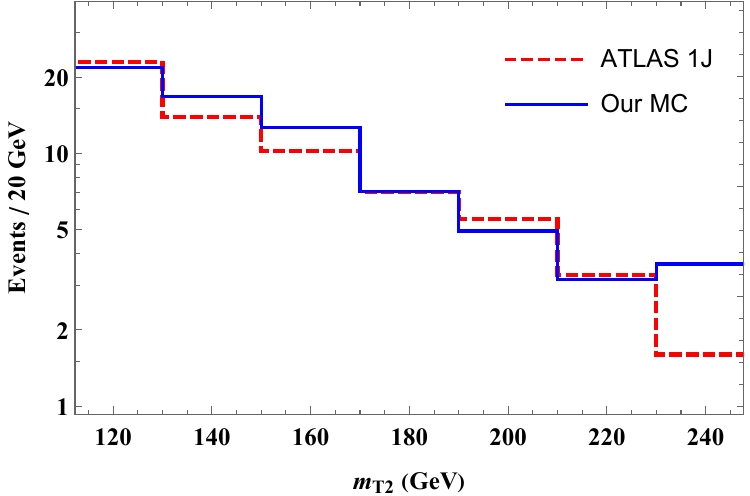}
         \caption{1J channel}
         \label{fig:slepton 1J MC}
     \end{subfigure}
\caption{Comparison between the ATLAS background prediction and our simulated diboson background from the slepton search~\cite{ATLAS:2019lff}.}
\label{fig:comparisonWWZZslepton}
\end{figure}

We also show comparison between simulated background events and ATLAS observed events in the slepton search~\cite{ATLAS:2019lff} for dominant $WW$ and $ZZ$ processes in Fig. \ref{fig:comparisonWWZZslepton}. We show both signal regions with 0 jet and 1 jet observed. There is a good agreement between our simulation and ATLAS estimates.

In order to calculate the bounds, we follow the strategy in Refs.~\cite{GAMBIT:2017qxg,Primulando:2020cbl}. We first calculate the likelihood of observing signal in a particular bin $i$, $s_i$, given by
\begin{equation}
     \mathcal{L}_{i}(n_i | s_i, b_i) =  \int_{0}^{\infty}\frac{(\xi (s_i + b_i ))^{n_i} \ e^{-\xi(s_i + b_i)}}{n_i !} P_i(\xi)  \,d\xi,  
\end{equation}
where $n_i$ and $b_i$ are the number of observed events and the predicted number of background events, respectively. $P_i(\xi)$ is probability defined by the log-normal function

\begin{equation}
    P_i(\xi) = \frac{1}{\sqrt{2\pi} \sigma_i} \frac{1}{\xi} \exp \left[{-\frac{1}{2}}\left(\frac{\ln \xi}{\sigma_i} \right) \right],
\end{equation}
here $\sigma_i$ is the relative uncertainties of the corresponding bin. We then determine the chi-square values, given by

\begin{equation}
    \chi^2 = -2 \sum_{i} (\ln \mathcal{L}(n_i | s_i, b_i)-\ln \mathcal{L}(n_i | s_i=0, b_i))
    \label{chisq}.
\end{equation}
This will facilitate computations related to the signal emanating from our model, allowing for a comparative analysis against the presently available data from LHC searches.

Now, armed with validated background simulations, we can proceed with signal simulations and explore the bounds on the dark coupling constant ($g_D$) using the chi-square approach with a 95\% C.L. requirement.

%Table 1 gives the cuts value for leptons, missing energy, jets, and other variables used.

%\begin{tabular}{ |p{3cm}|p{3cm}|p{3cm}|  }
%\multicolumn{2}{|c|}{} \\
%\hline
%& Cuts \\
%\hline
%$pT_{,e}$ & $>10$ GeV \\
%$|\eta_e|$ & $<2.47$  \\
%$pT_{,\mu}$ & $>10$ GeV  \\
%$|\eta_\mu|$    & $<2.7$ \\
%$pT_{j}$ & $>20$ GeV \\
%$|\eta_j|$ & $<2.4$    \\ 
%$\textbf{m_ll'}$ & $>100$ GeV  \\
%\hline
%\end{tabular}

%%%%%%%%%%%%%%%%%%%%%%%%%%%%%%%%%%%%%%%%%%%%%%%%%

\subsection{Results}
\label{sec:res}
Our search strategy is based on the darkstrahlung process, as illustrated in Fig. \ref{feynmandiag}, where the final state comprises a pair of OS leptons accompanied by missing energy. As mentioned before, our choice of benchmarks involved setting specific combinations of $m_\chi$ and $m_{Z_A}$ in accordance with the bounds derived from the monojet analysis~\cite{ATLAS:2021kxv}. Following~\cite{ATLAS:2021kxv}, we set the parameters of $g_Q$ and $g_{\chi}$ in Eq. (\ref{eq:monojet}) to be 0.25 and 1, respectively. This choice enables us to obtain constraints on the dark coupling constant ($g_D$) as a function of the $Z'$ mass. 

\begin{figure}[h]
\includegraphics[width=0.7\textwidth]{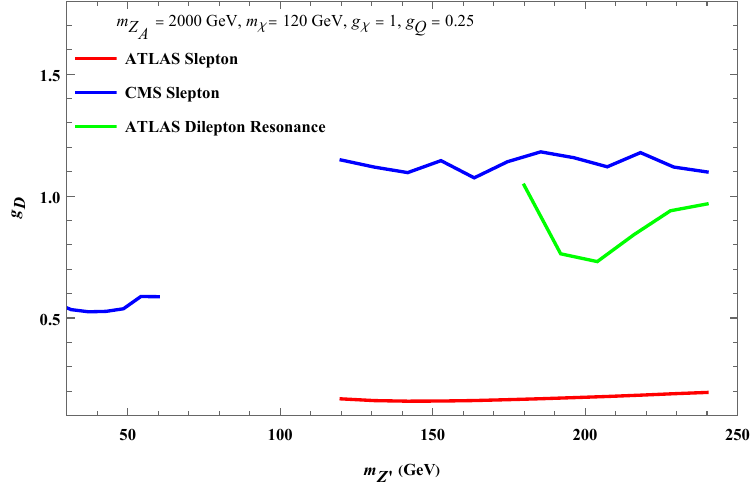}
\caption{The 95\% C.L. bounds on dark coupling constant ($g_D$) as a function of $m_{Z'}$ from recasting the ATLAS dilepton resonance search~\cite{ATLAS:2023tmv}, and ATLAS~\cite{ATLAS:2019lff} and CMS~\cite{CMS:2020bfa} slepton searches. In this figure we chose a benchmark of $m_{Z_A} = 2$ TeV and $m_\chi = 120$ GeV,  $g_Q = 0.25$ and $g_{\chi} = 1$, corresponding with the ATLAS monojet analysis~\cite{ATLAS:2021kxv}.}
\label{fig:bounds}
\end{figure}

\begin{figure}[h]
\includegraphics[width=0.7\textwidth]{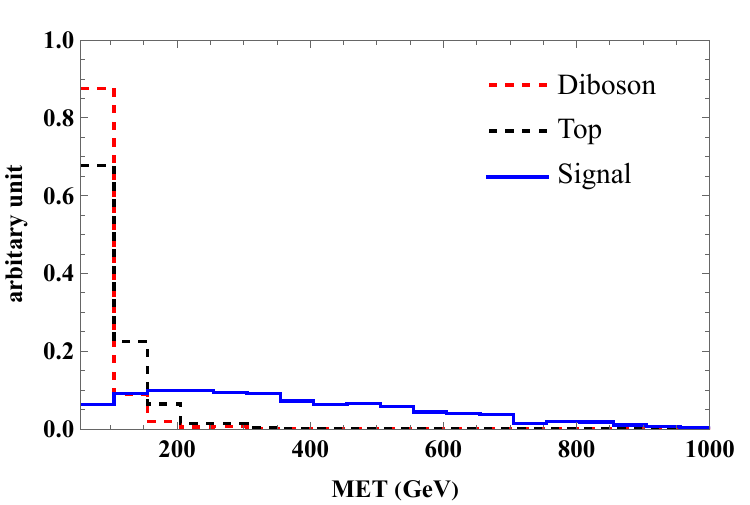}
\caption{The missing energy distribution of the signal and the main backgrouns of the ATLAS dilepton resonance search~\cite{ATLAS:2023tmv}. The benchmark for the signal is $m_{Z_A} = 2$ TeV, $m_\chi = 120$ GeV, $m_{Z'} = 200$ GeV,  $g_Q = 0.25$ and $g_{\chi} = 1$}.
\label{fig:metdist}
\end{figure}

The recasted bounds from ATLAS dilepton resonance search~\cite{ATLAS:2023tmv} as well as ATLAS~\cite{ATLAS:2019lff} and CMS~\cite{CMS:2020bfa} slepton searches for $m_{Z_A} = 2$ TeV and $m_\chi = 120$ GeV are shown in Fig. \ref{fig:bounds}. Comparing results from the ATLAS dilepton resonance search with the ATLAS slepton search, we can see that the ATLAS slepton search delivers stronger constraints on the darkstrahlung process. This observation may seem counterintuitive at first, as the ATLAS dilepton resonance search bins events based on the invariant mass of the dilepton, enabling detection of a dilepton resonance from the $Z'$ decay. However, the cut on the missing energy for the dilepton search is quite low ($E_T^\text{miss}$ > 55 GeV), while the ATLAS slepton search imposes a cut of  $E_T^\text{miss}$ > 110 GeV. From Fig.~\ref{fig:metdist}, we can see that the signal exhibits substantial missing energy. Hence the ATLAS slepton search can constrain more than the dilepton resonance search. This also suggests that the ATLAS dilepton resonance search could be improved by implementing a stronger cut on missing energy.

There is also a difference between the ATLAS and CMS slepton results for high values of $m_{Z'}$. It can be attributed to CMS having more search regions than ATLAS. The CMS signal region  includes the invariant mass of the two leptons to be 20 GeV < $m_{ll}$ < 65 GeV and $m_{ll}$ > 120 GeV regions, while ATLAS only looks in $m_{ll}$ > 121.2 GeV region for the same flavor leptons events. The CMS analysis, with its broader search coverage, tends to observe more background events, consequently leading to less restrictive bounds. Other consequences of the chosen signal regions are that only CMS can probe the case 20 GeV < $m_{Z'}$ < 65 GeV, while $m_{Z'}$ between 65 GeV and $\sim$120 GeV are unconstrained by any of the analyses. 

\begin{figure}
     \centering
     \begin{subfigure}[b]{0.45\textwidth}
         \centering
         \includegraphics[width=\textwidth]{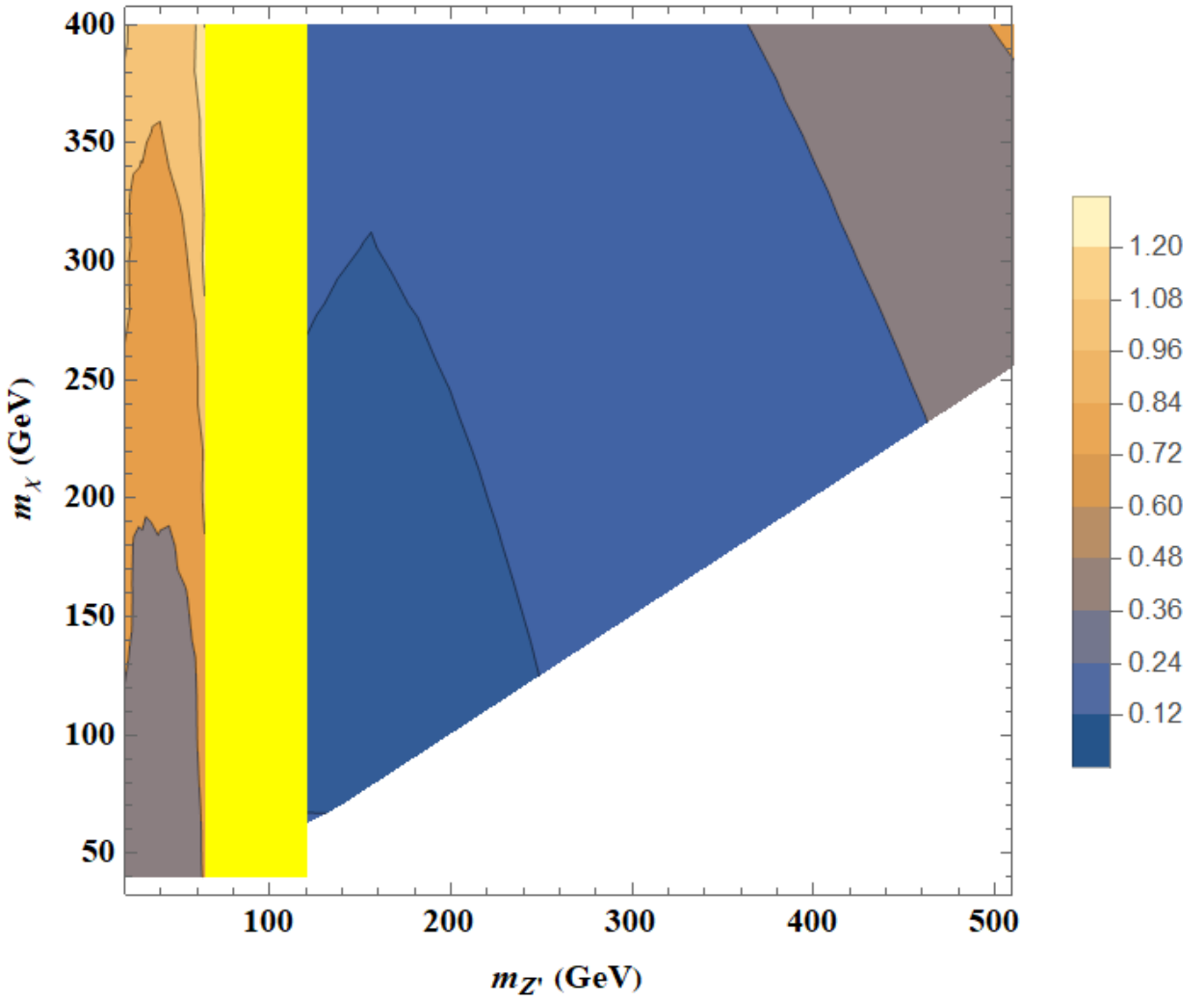}
         \caption{Unoptimized bounds}
         \label{fig:unoptimized3D}
     \end{subfigure}
     \hfill
     \begin{subfigure}[b]{0.45\textwidth}
         \centering
         \includegraphics[width=\textwidth]{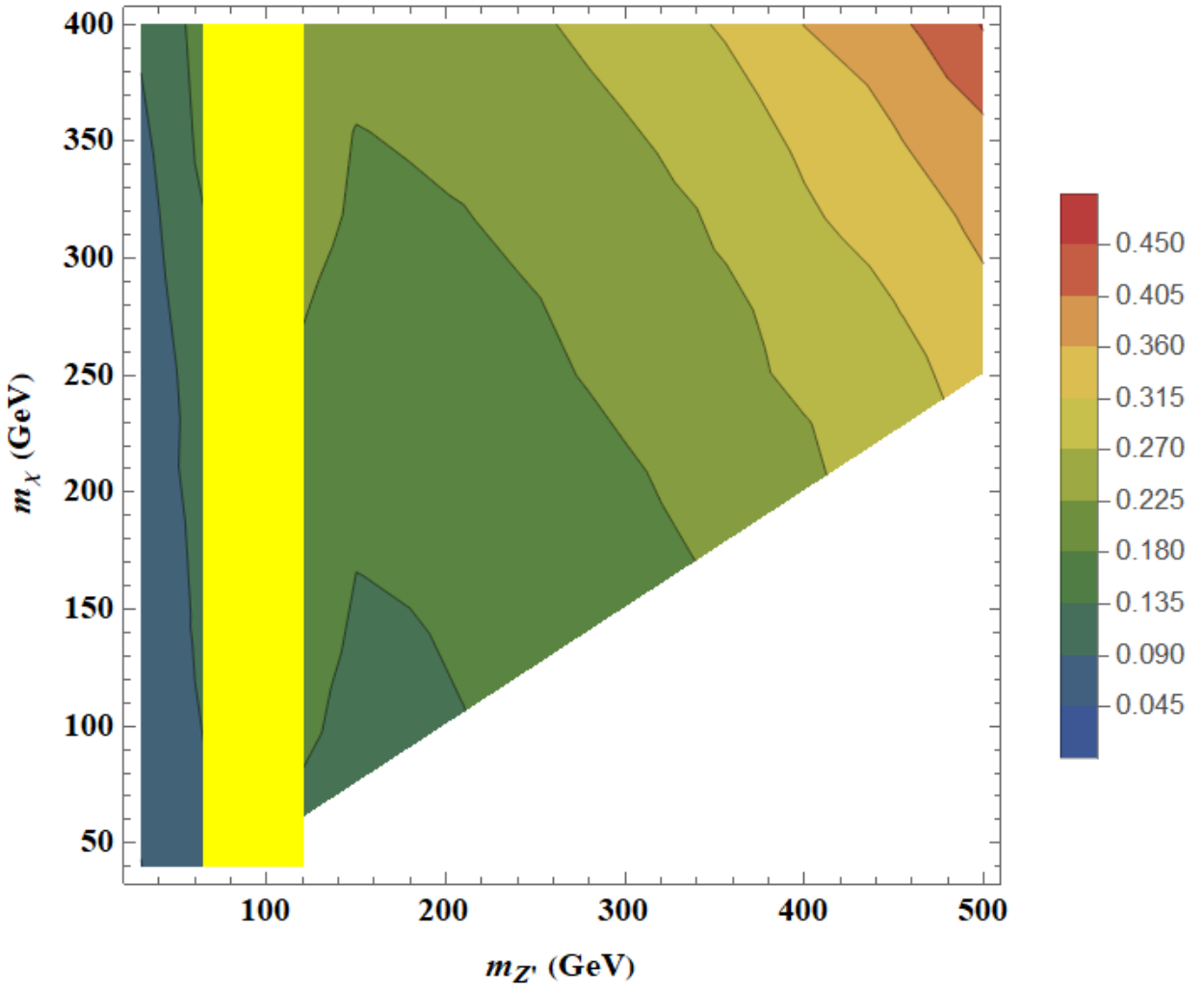}
         \caption{Optimized bounds}
         \label{fig:optimized3D}
     \end{subfigure}
        \label{fig:contour3D}
\caption{Constraints on $g_D$ at 95\% C.L. in the plane of $\chi$ and $Z'$ mass, from data presented in ATLAS~\cite{ATLAS:2019lff} and CMS~\cite{CMS:2020bfa} slepton searches (left figure). The right figure shows  projected reach of 95\% C.L. exclusion with optimized ATLAS dilepton resonance search~\cite{ATLAS:2023tmv}, by utilizing higher MET cut and dilepton invariant mass to $\pm 5$ GeV of $Z'$ mass}.
\end{figure}

Having explained the case for a particular value of $m_{\chi}$, we show the bounds on $g_D$ for various values of $m_{\chi}$ and $m_{Z'} $ in Fig.~\ref{fig:unoptimized3D}. The yellow region indicates an unexplored domain by both ATLAS and CMS slepton searches. The white region indicates the case which $m_{Z'} > 2 m_{\chi}$, in which the $Z'$ decays invisibly. In the region $20 \text{ GeV} < m_{Z'} < 65 \text{ GeV}$, we use the search region from CMS for comparison with our signal, while for $m_{Z'} > 121.2 \text{ GeV}$, the ATLAS search criteria are employed for the corresponding region. From the plot, one can see that for a fixed value of $m_{Z'}$, the heavier the dark matter mass is, the worse the bounds on $g_D$. This is because the signal production cross section becomes smaller as the final state masses get heavier.

\begin{figure}[h]
\includegraphics[width=0.7\textwidth]{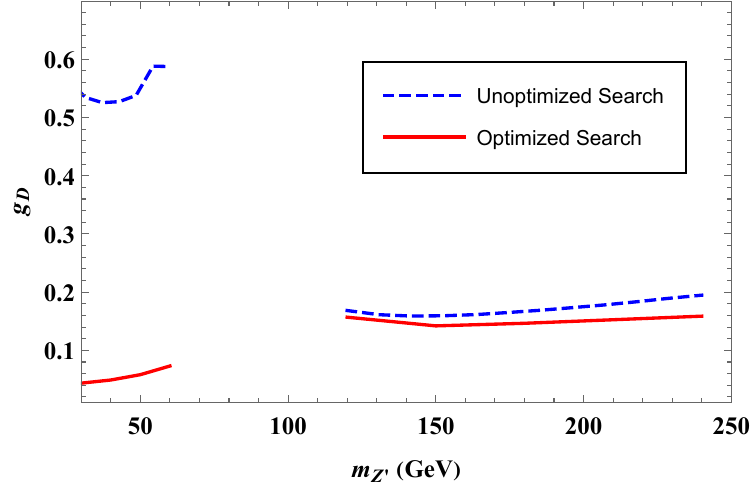}
\caption{Comparison between unoptimized and optimized search strategy for recasted ATLAS~\cite{ATLAS:2019lff} and CMS~\cite{CMS:2020bfa} slepton searches. Optimized search employs  dilepton invariant mass to $\pm 5$ GeV of the $Z'$ mass and stronger missing energy cuts from the ATLAS dilepton resonance search~\cite{ATLAS:2023tmv}. In this Figure we chose a benchmark of $m_{Z_A} = 2$ TeV and $m_\chi = 120$ GeV,  $g_Q = 0.25$ and $g_{\chi} = 1$, corresponding with the ATLAS monojet analysis~\cite{ATLAS:2021kxv}.}
\label{fig:opt2D}
\end{figure}

As we mentioned before, the ATLAS dilepton resonance search~\cite{ATLAS:2023tmv} can be improved to search for the darkstrahlung process by simply increasing the missing energy. We simulate an optimized search by performing simulations for signal and background with the ATLAS cuts for dilepton resonance search with additional cuts on the invariant mass of the dilepton $\left| m_{ll} - m_{Z'}\right| < 5$ GeV and missing energy $E_T^\text{miss} > 250$ GeV. The improvements can be seen in Fig.~\ref{fig:opt2D}, where the red lines show the expected constraints from optimized search and the dashed blue lines show the bounds from recasted slepton search from Fig. \ref{fig:bounds}. For the recasted bounds with $m_{Z'} > 120$ GeV, we only show the ATLAS slepton search results.

%Since the ATLAS and CMS signal regions are optimized to search for slepton, the above constrains can be improved. In particular, it is expected that the invariant mass of the two leptons coming from the $Z'$ decay should be clustered around $m_{Z'}$. Therefore an optimized search would employ a narrower invariant mass bins. We simulate an optimized search by performing simulations for signal and background with the ATLAS cuts for slepton search with an additional cut $\left| m_{ll} - m_{Z'}\right| < 5$ GeV. The improvements can be seen in Fig. \ref{fig:opt2D}, where the blue lines show the expected constraints from optimized search and the red lines shows the bounds from recasted slepton search from Fig. \ref{fig:bounds}. For the recasted bounds with $m_{Z'} \gtrsim 120$ GeV, we only show the ATLAS results. 

From the figure we see that for $m_{Z'} < 65$ GeV, there is a significant improvement. This is due to the original CMS slepton search signal region that includes both  20 GeV $< m_{ll} < $ 65 GeV and $m_{ll} > $120 GeV. Hence having a tighter cut on $m_{ll}$ helps to reduce the SM background significantly. For higher values of $m_{Z'}$ there is a factor of 18\% improvements compared with the recasted ATLAS slepton search. By comparison, there is a factor of six improvements from the ATLAS dilepton resonance search space. Further enhancements might be achieved if the LHC experiments segregate the signal regions based on $E_T^\text{miss}$ significance in addition to imposing stronger cuts on missing energy. The projected results for the optimized search also result in stronger bounds on $g_D$ for various values of $m_{\chi}$ and $m_{Z'} $, as shown in Fig.~\ref{fig:optimized3D}. There is an improvement by a factor of up to 6 in the light $m_{Z'}$ region, and up to 25\% in the heavy $m_{Z'}$ region.  

%\begin{figure}[h]
%\includegraphics[width=0.7\textwidth]{OptvsNoOpt2D_diboson.pdf}
%\caption{Comparison between unoptimized and optimized search strategy for recasted ATLAS dilepton resonance search~\cite{ATLAS:2023tmv}. Optimized search employs narrowed dilepton invariant mass to $\pm 5$ GeV of $Z'$ mass. In this Figure we chose a benchmark of $m_{Z_A} = 2$ TeV and $m_\chi = 120$ GeV,  $g_Q = 0.25$ and $g_{\chi} = 1$, corresponding with the ATLAS monojet analysis~\cite{ATLAS:2021kxv}.}
%\label{fig:opt2D_dilepton}
%\end{figure}

%Such usage of tighter cuts show that the projected results for the search will have better values, with improvement shown in Figure~\ref{fig:optimized3D}. In doing so, we employed ATLAS search and events for both 20 GeV < $m_{Z'}$ < 65 GeV and  $m_{Z'}$ > 121.2 region. We argue that based on Figure~\ref{fig:bounds}, ATLAS will give better bounds than CMS. It is reasonable to also use it in lower mass search region. To do so, we did simulation of ATLAS search in lower mass search in order to obtain the expected events, and calculate the bounds for $g_D$ afterwards. We use narrowed dilepton invariant mass to $\pm 5$ GeV of $Z'$ mass in our calculation, and impose no MET cut. \cblue{Did we really???} 

%The improvement in our optimized search can be seen more clearly in Fig.~\ref{fig:opt2D}. It can be seen that for low values of $m_{Z'}$, the improvement can reach to a factor of 10. In the case of higher values of $m_{Z'}$, we have less improvement in the optimized search, only by a factor of 2. 

%%%%%%%%%%%%%%%%%%%%%%%%%%%%%%%%%%%%%%%%%
\section{Long-Lived $Z'$}
\label{sec:LLP}

\begin{figure}[h]
\includegraphics[width=0.7\textwidth]{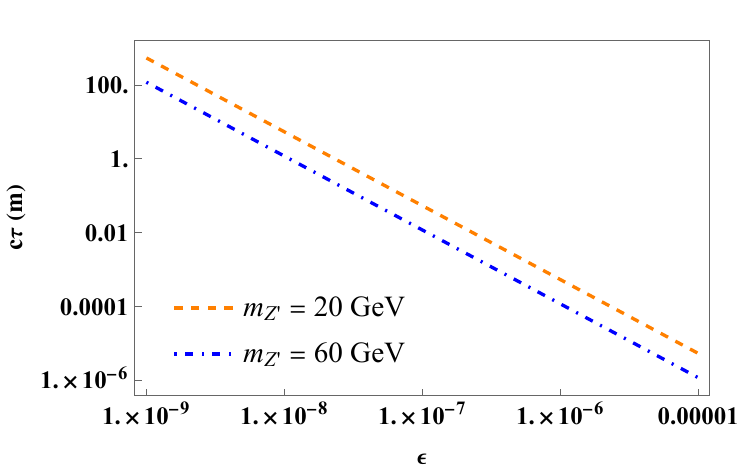}
\caption{The decay length ($c\tau$) as a function of $\epsilon$ for $m_{Z'} = 20$ and 60 GeV.}
\label{fig:ctau}
\end{figure}

In the previous section, we derived constraints for the promptly decaying $Z'$. The bounds for that case are independent of $\epsilon$ as long as the DM decays only to SM particles through its mixing.  However as the value of $\epsilon$ gets smaller, the lifetime of the $Z'$ gets longer, leading to a displaced decay of the boson, illustrated in Fig. \ref{fig:ctau}.

In this section, we will discuss the LHC constraints on long-lived $Z'$ produced by the darkstrahlung process. We specifically focus on the CMS search targeting long-lived particles decaying into a pair of muons \cite{CMS:2022qej}. This search utilizes data from the 2016 and 2018 runs, amounting to 36.3 fb$^{-1}$ and 61.3 fb$^{-1}$, respectively, both at a center-of-mass energy of $\sqrt{s} = 13$ TeV. The reconstructed muon pairs are categorized based on whether both muons are reconstructed in the tracker and muon system (TMS-TMS categories), both muons are reconstructed only in the muon system (STA-STA categories), or one muon is reconstructed in the tracker and muon system while the other is solely reconstructed in the muon system (STA-TMS categories).

For each event a primary vertex (PV) is defined as the vertex corresponding to the hardest scattering in the event. Since the muon pair comes from long lived $Z'$, it is expected that the reconstructed common vertex (CV) of the muon will have some displacement with respect to the PV. The CMS collaboration introduces the transverse decay length, $L_{xy}$, representing the distance of the PV and CV in the transverse plane. Additionally, the transverse decay parameter for each reconstructed muon, $d_0$, is defined as the distance of the closest approach of the muon track relative to the PV in the transverse plane.

The CMS collaboration provides a procedure to reinterpret their results, presenting the efficiency maps for the minimum $p_T$ of the generated muons, minimum $d_0$ of generated muons and $L_{xy}$. These maps are categorized for each event category and for the 2016 and 2018 runs. Furthermore, the collaboration supplies signal regions for benchmark $Z'$ masses, along with corresponding estimated backgrounds and observed event counts.

\begin{figure}
     \centering
     \begin{subfigure}[b]{0.45\textwidth}
         \centering
         \includegraphics[width=\textwidth]{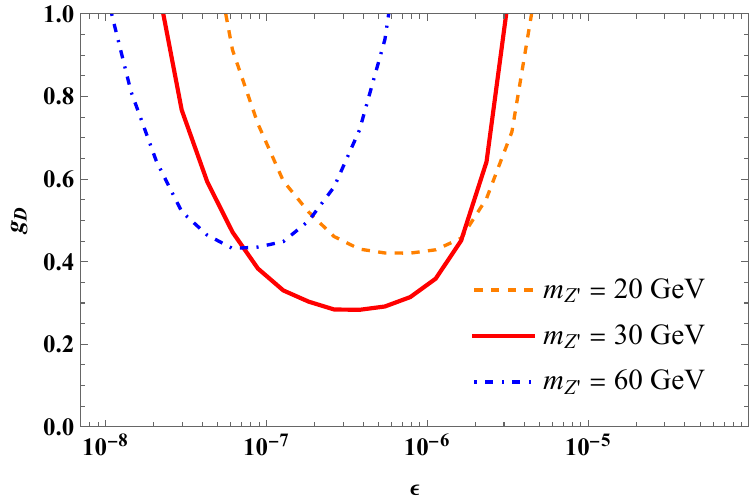}
         \caption{$m_{\chi} = 120$ GeV.}
         \label{fig:long122}
     \end{subfigure}
     \hfill
     \begin{subfigure}[b]{0.45\textwidth}
         \centering
         \includegraphics[width=\textwidth]{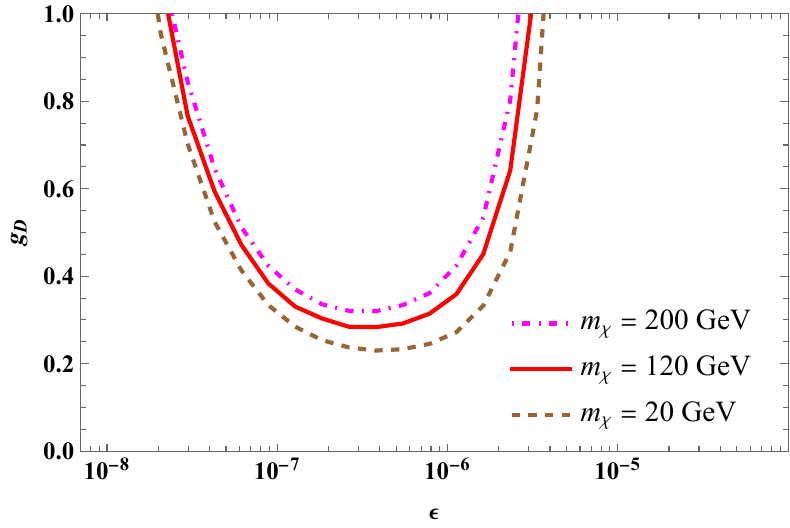}
         \caption{$m_{Z'} = 30$ GeV}
         \label{fig:long30}
     \end{subfigure}
        \label{fig:smallepsilon}
    \caption{The 95\% C.L. constrains for $g_D$ as a function of $\epsilon$. For the left figure, $m_\chi$ is fixed while $m_{Z'}$ is varied. $m_{Z'}$ is fixed and $m_\chi$ is varied for the right figure.}
\end{figure}

We simulate the $Z'$ darkstrahlung process using Madgraph5 \cite{Alwall:2014hca} for a specific set of parameters of $m_{Z'}$, $g_D$, $m_{Z_A}$, $g_{\chi}$ and $g_{Q}$. Subsequently, we simulate the displaced decay of $Z'$ and calculate the values of $L_{xy}$ and $d_0$ for each event. The expected number of events are then calculated using the efficiency maps and the constraints can be derived from it. The results for $m_\chi = 120$ GeV and various values of $m_{Z'}$ are presented in Fig. \ref{fig:long122}. The searches are most sensitive for $m_{Z'}$ around 30 GeV, with lower sensitivity for higher $m_{Z'}$ values due to lower cross sections and diminished sensitivity for lower $m_{Z'}$ values owing to increased background contributions. The constraints for the benchmark value $m_{Z'} = 150$ GeV is already out of the range of the plot. Since this search is sensitive for displaced muon pair decaying in either tracker or muon system, it cannot constrain $\epsilon \gtrsim 10^{-5}$. In this case the constraints from Section \ref{sec:res} apply. This search is also less sensitive for $\epsilon \lesssim 10^{-8}$ in which the $Z'$ decays outside the muon system. The dominant signal for this low value of $\epsilon$ would be the monojet or a more dedicated lifetime frontier detectors~\cite{Feng:2017uoz,Curtin:2018mvb}. 

In Fig. \ref{fig:long30} we also show the case when the value of $m_{Z'}$ kept fixed while varying $m_{\chi}$. In this scenario, the constraints are primarily dictated by the cross section; specifically, as the mass of the dark matter particle, $m_\chi$, increases, the constraints become weaker.

%\begin{figure}[h]
%\includegraphics[width=0.7\textwidth]{Long122.pdf}
%\caption{The 95\% CL constrains for $g_D$ in the case of $m_\chi = 120$ GeV and $m_{Z_A} = 2.1$ TeV for various values of $m_{Z'}$}
%\label{fig:long122}
%\end{figure}

%\begin{figure}[h]
%\includegraphics[width=0.7\textwidth]{Long30.pdf}
%\caption{The 95\% CL constrains for $g_D$ in the case of $m_{Z'} = 30$ GeV and $m_{Z_A} = 2.1$ TeV for various values of $m_{\chi}$}
%\label{fig:long30}
%\end{figure}

In all discussions above, we have assumed no mixing between $Z_A$ and $Z'$. However, this mixing can be introduced from a loop of $\chi$. The size of the mixing is given by $\varepsilon \sim \frac{g_\chi g_D}{16 \pi^2} \log\left(\frac{m_\chi}{\Lambda}\right)$, where $\Lambda$ is the scale of the UV completion~\cite{Holdom:1985ag} of this simplified model. Depending on the scale of the completion, this mixing can get to be as large as $\mathcal{O} \left(10^{-3}\right)$. The mixing will introduce a coupling between the $Z'$ and the quarks in the order of $g_{Z'-Q} \sim \varepsilon g_Q \frac{m_{Z'}}{m_{Z_A}}$, which can get as large as $\mathcal{O} \left(10^{-5}\right)$. This large value of $g_{Z'-Q}$ makes the decay of $Z'$ to be prompt and the bounds from the long-lived $Z'$ to be not applicable. More relevant bounds coming from the CMS search for prompt dijet resonance associated with a photon will be more relevant~\cite{CMS:2019xai}. However, this search still can not constrain the value of $g_{Z'-Q} \sim \mathcal{O}(10^{-5})$.% and the UA1 search on dilepton resonance~\cite{UA1:1988rck}. Neither of the searches can constrain $g_{Z'-Q} \sim \mathcal{O}(10^-5)$ and $\epsilon \lesssim 10^{-5}$.

%%%%%%%%%%%%%%%%%%%%%%%%%%%%%%%%%%%%%%%%%
\section{Conclusion}
\label{sec:conc}
In this work, we discussed LHC constraints on a model where the dark sector consists of a dark matter charged under a new hidden $U(1)$ symmetry. The dark matter is produced via a heavy $Z_A$ mediator and is constrained by the LHC monojet bounds. The produced dark matter can radiate the gauge boson of the hidden $U(1)$ symmetry, which we call the $Z'$. We consider the case where the $Z'$ can only kinematically decay to the SM particles, leading to signatures of a lepton pair associated with missing energy.

We recasted the LHC searches to constrain the model. We found that the LHC run 2 can significantly constrain the model. There are already several searches by ATLAS and CMS that can be utilized to constrain the model. ATLAS dilepton resonance search, which bins their events into dilepton invariant mass, should be able to detect the resonance coming from the decay of $Z'$, with relatively low background. ATLAS and CMS search for slepton which looks for a dilepton pair with accompanying missing energy, can also be used as a comparison. For the masses of the dark matter and mediator at around the monojet exclusion, we have found that the recasted  searches can exclude the $\mathcal{O}(100)$ GeV $Z'$ with the dark gauge coupling, $g_D$, larger than 1.2. While the dilepton resonance search is preferred due to the sensitivity to $Z'$ decay, we found that the obtained bounds are not optimum due to the low missing energy cut related to their chosen benchmark model. By simulating a higher cut on missing energy, the bounds will improve by a factor of 6.

%For the masses of the dark matter and mediator at around the monojet exclusion, we have found that the recasted slepton search can exclude the $\mathcal{O}(100)$ GeV $Z'$ with the dark gauge coupling, $g_D$, smaller than 1. A more dedicated search for this channel will improve the bounds substantially. As an example for this point, we simulated an additional cut in the lepton pair invariant mass corresponding with the searched $Z'$ mass. We found that the bounds get better, even at one order of magnitude for lighter $Z'$.

We also consider the LHC bound for cases in which the $Z'$ is long-lived due to the low value of $\epsilon$ by recasting a CMS search of displaced di-muon pairs. We found that the search is sensitive for $m_{Z'} \lesssim 100$ GeV and $10^{-8} \lesssim \epsilon \lesssim 10^{-5}$, assuming negligible mixing between $Z'$ and $Z_A$. For higher values of $\epsilon$, the prompt search is more relevant, while for monojet search can constrain lower values of $\epsilon$.

We have shown that the current LHC Run-2 searches are capable of ruling out a large region of hidden $ Z'$'s parameter space. We recommend that for the incoming runs, both ATLAS and CMS collaborations should employ a tighter condition by imposing a higher MET cut. This should improve the bounds on parameter space, as shown in this work.

%In this work, we discussed LHC constrains on model where dark sector consist of a dark matter charged under a new-hidden $U(1)$ symmetry. This will allow the coupling of DM to SM particles, and provides possibility of a distinctive collider signature. This signature is dubbed darkstrahlung process, with final state comprises of dilepton resonances as a decay product of $Z'$, and missing transverse energy. The $Z'$ is radiated from dark matter particle. We performed an optimization to get improved bounds for such model by recasting the slepton search for two non resonant OS leptons with missing energy. These searches  was done previously by ATLAS and CMS. We show that by using tighter cuts for invariant mass bins, we are able to get improvement of a factor of ten in light mass region and a factor of two in higher mass regions.   

%%%%%%%%%%%%%%%%%%%%%%%%%%%%%%%%%%%%%%%%%
\begin{acknowledgments}
 A.\,F. was supported by funding from Higher Education Funding Center (BPPT) Kemendikbudristek and Indonesia Endowment Fund for Education Agency (LPDP) under funding ID 202209092458. R.\,P. and Q.\,M.\,B.\,S was supported by Direktorat Riset, Teknologi dan Pengabdian kepada Masyarakat Direktorat Jenderal Pendidikan Tinggi, Riset, dan Teknologi, Kementerian Pendidikan, Kebudayaan, Riset, dan Teknologi Republik
Indonesia in the year 2024 with contract number III/LPPM/2024-06/110-PE and 006/SP2H/RT-MONO/LL4/2024.\, B.\,E.\,G. was supported by PPMI KK FMIPA ITB 2024. 

\end{acknowledgments}

\appendix
\section{The ratio of $\frac{\sigma_{q \bar q \rightarrow \chi \bar\chi Z'}}{\sigma_{q \bar q \rightarrow \chi \bar\chi}}$}
\label{sec:app}

To calculate the value of $\frac{\sigma_{q \bar q \rightarrow \chi \bar\chi Z'}}{\sigma_{q \bar q \rightarrow \chi \bar\chi}}$, we will assume productions of onshell heavy $Z_A$:
\begin{equation}
\begin{split}
    \frac{\sigma_{q \bar q \rightarrow \chi \bar\chi Z'}}{\sigma_{q \bar q \rightarrow \chi \bar\chi}} &= \frac{\sigma_{q \bar q \rightarrow Z_A}\text{BR}_{Z_A \rightarrow \chi \bar\chi Z'}}{\sigma_{q \bar q \rightarrow Z_A}\text{BR}_{Z_A \rightarrow \chi \bar\chi}} \\
    &= \frac{\Gamma_{Z_A \rightarrow \chi \bar\chi Z'}}{\Gamma_{Z_A \rightarrow \chi \bar\chi}}. \\
\end{split}    
\end{equation}

\begin{figure}
     \centering
     \begin{subfigure}[b]{0.45\textwidth}
         \centering
         \begin{tikzpicture}
			\begin{feynman}
				\vertex (c){$Z_A$};
				\vertex [right=of c] (d);
				\vertex [above right=of d] (e);
				\vertex [above right=of e] (f){$\chi$};
				\vertex [right=of e] (g){$Z'$};
				\vertex [below right=of d] (j);
				\vertex [below right=of j] (k){$\bar{\chi}$};
				
				\diagram{
					(c) -- [boson, momentum=$k$] (d);
					(d) -- [fermion, edge label=$\chi$] (e) -- [fermion,momentum=$p_1$] (f);
					(d) -- (j);
                    (k) -- [fermion] (d);
                    (j) --[momentum'=$p_2$](k);
					(e) -- [boson,momentum'=$p_3$] (g);
				};
			\end{feynman}
		\end{tikzpicture}
  \caption{}
         \label{fig:fey1}
     \end{subfigure}
     \hfill
     \begin{subfigure}[b]{0.45\textwidth}
         \centering
         \begin{tikzpicture}
			\begin{feynman}
				\vertex (c){$Z_A$};
				\vertex [right=of c] (d);
				\vertex [above right=of d] (e);
				\vertex [above right=of e] (f){$\chi$};
				\vertex [below right=of d] (j);
				\vertex [below right=of j] (k){$\bar{\chi}$};
                \vertex [right=of j] (g){$Z'$};

				\diagram{
					(c) -- [boson, momentum=$k$] (d);
					(d) -- [fermion] (f);
                    (e) --[momentum=$p_1$](f);
					(j) -- [fermion, edge label=$\bar\chi$](d);
                    (k) -- [fermion] (j);
                    (j) --[momentum'=$p_2$](k);
					(j) -- [boson,momentum=$p_3$] (g);
				};
			\end{feynman}
		\end{tikzpicture}
  \caption{}
         \label{fig:fey2}
     \end{subfigure}
    \caption{Feynman diagrams for the decay of $Z_A \rightarrow \chi\bar\chi Z'$.}
    \label{feydec}
\end{figure}

The decay width of $Z_A \rightarrow \chi \bar\chi$ is given by
\begin{equation}
    \Gamma_{Z_A \rightarrow \chi \bar\chi} = \frac{g_X^2 m_{Z_A}}{12\pi} \left(1-\frac{4m_\chi^2}{m_{Z_A}^2}\right)^{3/2}.
\end{equation}
The decay width of $Z_A \rightarrow \chi \bar\chi Z'$ is more complicated since it involves three particles in the final state. The Feynman diagrams for this decay are shown in Fig.~\ref{feydec}. The amplitude of $Z_A \rightarrow \chi \bar\chi Z'$ is given by
\begin{equation}
    \mathcal M = g_X g_D \bar u_1 \left(\gamma^\alpha \frac{\slashed{p}_1 + \slashed{p}_3 + m_\chi}{\left(p_1 + p_3\right)^2 - m_\chi^2} \gamma^\mu \gamma^5 + \gamma^\mu \gamma^5 \frac{-\slashed{p}_2 - \slashed{p}_3 + m_\chi}{\left(p_2 + p_3\right)^2 - m_\chi^2} \gamma^\alpha\right) v_2 \, \epsilon_{k,\mu} \epsilon^*_{3,\alpha}.
\end{equation}

The squared amplitude averaged over the spin is given by
\begin{equation}
    \langle \left| \mathcal M \right|^2 \rangle = g_X^2 g_D^2 \frac{\sum_{i,j} d_{i,j} \left(m_{12}^2\right)^i \left(m_{23}^2\right)^j }{D},
\end{equation}
where $m_{12}^2 = (p_1 + p_2)^2$ and $m_{23}^2 = (p_2 + p_3)^2$. 
The coefficients of the numerator are given by
\begin{equation}
    \begin{split}
        d_{0,0} =& -8 \left(2 m_\chi^6 \left(3 m_{Z_A}^4+ m_{Z'}^4\right)+2 m_\chi^8 m_{Z_A}^2+m_{Z_A}^4 m_{Z'}^2 \left(m_{Z_A}^2+m_{Z'}^2\right)^2-m_\chi^4 \left(m_{Z_A}^2+2 m_{Z'}^2\right) \left(3 m_{Z_A}^4+4 m_{Z_A}^2 m_{Z'}^2-m_{Z'}^4\right)\right. \\ & \left.+m_\chi^2 m_{Z_A}^2 \left(m_{Z_A}^2+m_{Z'}^2\right) \left(3 m_{Z_A}^4+2 m_{Z_A}^2 m_{Z'}^2-3 m_{Z'}^4\right)\right), \\
        d_{0,1} =& 8 \left(2 m_\chi^2+m_{Z_A}^2+m_{Z'}^2\right) \left(m_{Z_A}^6+m_{Z_A}^4 \left(6 m_\chi^2+4 m_{Z'}^2\right)+m_{Z_A}^2 \left(4 m_\chi^4+m_{Z'}^4\right)+2 m_\chi^2 m_{Z'}^4\right), \\
        d_{0,2}=& -8 \left(3 m_{Z_A}^6+2 m_{Z_A}^4 \left(7 m_\chi^2+4 m_{Z'}^2\right)+m_{Z_A}^2 \left(12 m_\chi^4+3 m_{Z'}^4+8 m_\chi^2 m_{Z'}^2\right)+2m_\chi^2 m_{Z'}^4\right), \\
        d_{0,3}=& 32 m_{Z_A}^2 \left(2 m_\chi^2+m_{Z_A}^2+m_{Z'}^2\right), \\
        d_{0,4}=& -16 m_{Z_A}^2, \\
        d_{1,0} =& 8 \left(m_{Z_A}^6 \left(5 m_\chi^2+2 m_{Z'}^2\right)+2 m_{Z_A}^4 \left(m_\chi^4+m_{Z'}^4\right)+m_{Z_A}^2 \left(16 m_\chi^6-7 m_\chi^2 m_{Z'}^4\right)+6 m_\chi^4 m_{Z'}^4+4 m_\chi^6 m_{Z'}^2\right), \\
        d_{1,1} =& -8 \left(m_{Z_A}^6+m_{Z_A}^4 \left(22 m_\chi^2+4 m_{Z'}^2\right)+m_{Z_A}^2 \left(36 m_\chi^4+m_{Z'}^4+20 m_\chi^2 m_{Z'}^2\right)+6
   m_\chi^2 m_{Z'}^4+8 m_\chi^4 m_{Z'}^2\right), \\
   d_{1,2} =& 32 \left(m_{Z_A}^4+m_{Z_A}^2 \left(6 m_\chi^2+m_{Z'}^2\right)+m_\chi^2 m_{Z'}^2\right), \\
   d_{1,3} =& -32 m_{Z_A}^2, \\
   d_{1,4} = & 0 ,\\
   d_{2,0} =& -8 \left(m_{Z_A}^4 \left(3 m_\chi^2+m_{Z'}^2\right)+m_{Z_A}^2 \left(9 m_\chi^4-3 m_\chi^2 m_{Z'}^2\right)+2 m_\chi^4 \left(\text{m$\chi
   $}^2+3 m_{Z'}^2\right)\right), \\
   d_{2,1} =& 8 \left(4 m_\chi^4+m_{Z_A}^4+m_{Z_A}^2 \left(20 m_\chi^2+m_{Z'}^2\right)+6 m_\chi^2 m_{Z'}^2\right), \\ 
   d_{2,2} =& -8 \left(2 m_\chi^2+3 m_{Z_A}^2\right), \\
   d_{2,3} =& d_{2,4} = 0, \\
   d_{3,0} =& 8 m_\chi^2 \left(2 m_\chi^2+m_{Z_A}^2\right), \\
   d_{3,1} =& -8 \left(2 m_\chi^2+m_{Z_A}^2\right), \\
   d_{3,2} =& d_{3,3} = d_{3,4} = 0, 
    \end{split}
\end{equation}
and the denominator is given by 
\begin{equation}
    D = 3 m_{Z_A}^2 \left(m_{23}^2-m_\chi^2\right)^2 \left(-m_{12}^2-m_{23}^2+m_\chi^2+m_{Z_A}^2+m_{Z'}^2\right)^2.
\end{equation}

\begin{figure}[h]
\includegraphics[width=0.7\textwidth]{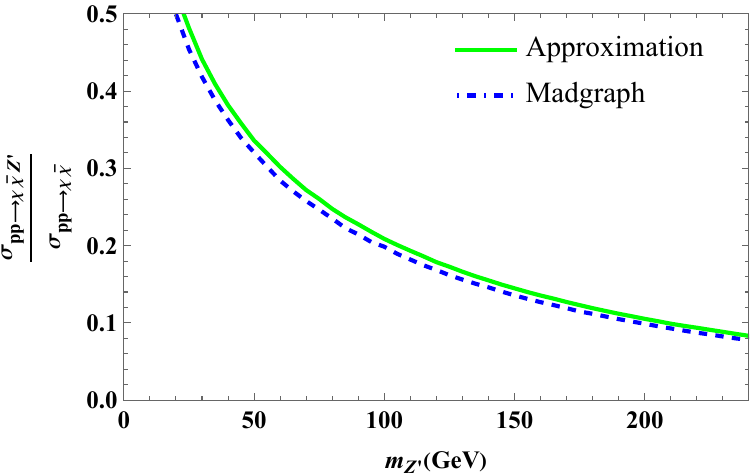}
\caption{The ratio $\frac{\sigma_{p \bar p \rightarrow \chi \bar\chi Z'}}{\sigma_{p \bar p \rightarrow \chi \bar\chi}}$ as a function of $m_{Z'}$ from Madgraph results and our approximation from $\frac{\sigma_{u \bar u \rightarrow \chi \bar\chi Z'}}{\sigma_{u \bar u \rightarrow \chi \bar\chi}}$.}
\label{fig:approx}
\end{figure}

The decay width is then given by
\begin{equation}
\label{int}
    \Gamma_{Z_A \rightarrow \chi\bar\chi Z'} = \iint \frac{1}{\left(2\pi\right)^3} \frac{1}{32 m_{Z_A}^3}\langle \left| \mathcal M \right|^2 \rangle dm_{12}^2 dm_{23}^2,
\end{equation} 
where the limits of integration are
\begin{equation}
    \begin{split}
        \left(m_{23}\right)_\text{max} &= \left(E_2^* + E_3^* \right)^2 - \left( \sqrt{E_2^{*2} - m_\chi^2} - \sqrt{E_3^{*2} - m_{Z'}^2} \right)^2, \\
        \left(m_{23}\right)_\text{min} &= \left(E_2^* + E_3^* \right)^2 - \left( \sqrt{E_2^{*2} - m_\chi^2} + \sqrt{E_3^{*2} - m_{Z'}^2} \right)^2, \\
        \left(m_{12}\right)_\text{min} &= 4 m_\chi^2, \\
        \left(m_{12}\right)_\text{max} &= \left( m_{Z_A} - m_{Z'}\right)^2,
    \end{split}
\end{equation}
with 
\begin{equation}
    \begin{split}
        E_2^* &= \frac{m_{12}}{2}, \\
        E_3^* &= \frac{m_{Z_A}^2 - m_{12}^2 - m_{Z'}^2}{2 m_{12}}.
    \end{split}
\end{equation}
The integration can not be done analytically. However, the value of $(m_{12})_\text{max}$ gets smaller as the $Z'$ gets heavier and the phase space closes. This could explain the dependence on the $m_{Z'}$ in the cross-section ratio. We can numerically integrate Eq.(\ref{int}) with the results shown in Fig.\ref{fig:approx}. 

%%%%%%%%%%%%%%%%%%%%%%%%%%%%%%%%%%%%%%%%%
%\bibliographystyle{unsrt}
\bibliography{paper_v1} 

\begin{thebibliography}{50}%
\makeatletter
\providecommand \@ifxundefined [1]{%
 \@ifx{#1\undefined}
}%
\providecommand \@ifnum [1]{%
 \ifnum #1\expandafter \@firstoftwo
 \else \expandafter \@secondoftwo
 \fi
}%
\providecommand \@ifx [1]{%
 \ifx #1\expandafter \@firstoftwo
 \else \expandafter \@secondoftwo
 \fi
}%
\providecommand \natexlab [1]{#1}%
\providecommand \enquote  [1]{``#1''}%
\providecommand \bibnamefont  [1]{#1}%
\providecommand \bibfnamefont [1]{#1}%
\providecommand \citenamefont [1]{#1}%
\providecommand \href@noop [0]{\@secondoftwo}%
\providecommand \href [0]{\begingroup \@sanitize@url \@href}%
\providecommand \@href[1]{\@@startlink{#1}\@@href}%
\providecommand \@@href[1]{\endgroup#1\@@endlink}%
\providecommand \@sanitize@url [0]{\catcode `\\12\catcode `\$12\catcode
  `\&12\catcode `\#12\catcode `\^12\catcode `\_12\catcode `\%12\relax}%
\providecommand \@@startlink[1]{}%
\providecommand \@@endlink[0]{}%
\providecommand \url  [0]{\begingroup\@sanitize@url \@url }%
\providecommand \@url [1]{\endgroup\@href {#1}{\urlprefix }}%
\providecommand \urlprefix  [0]{URL }%
\providecommand \Eprint [0]{\href }%
\providecommand \doibase [0]{https://doi.org/}%
\providecommand \selectlanguage [0]{\@gobble}%
\providecommand \bibinfo  [0]{\@secondoftwo}%
\providecommand \bibfield  [0]{\@secondoftwo}%
\providecommand \translation [1]{[#1]}%
\providecommand \BibitemOpen [0]{}%
\providecommand \bibitemStop [0]{}%
\providecommand \bibitemNoStop [0]{.\EOS\space}%
\providecommand \EOS [0]{\spacefactor3000\relax}%
\providecommand \BibitemShut  [1]{\csname bibitem#1\endcsname}%
\let\auto@bib@innerbib\@empty
%</preamble>
\bibitem [{\citenamefont {Albert}\ \emph {et~al.}()\citenamefont {Albert} \emph
  {et~al.}}]{Albert:2022xla}%
  \BibitemOpen
  \bibfield  {author} {\bibinfo {author} {\bibfnamefont {A.}~\bibnamefont
  {Albert}} \emph {et~al.},\ }\href@noop {} {\bibinfo {title} {{Displaying dark
  matter constraints from colliders with varying simplified model
  parameters}}},\ \bibinfo {note} {arXiv:2203.12035}\BibitemShut {NoStop}%
\bibitem [{\citenamefont {Buchmueller}\ \emph {et~al.}(2015)\citenamefont
  {Buchmueller}, \citenamefont {Dolan}, \citenamefont {Malik},\ and\
  \citenamefont {McCabe}}]{Buchmueller:2014yoa}%
  \BibitemOpen
  \bibfield  {author} {\bibinfo {author} {\bibfnamefont {O.}~\bibnamefont
  {Buchmueller}}, \bibinfo {author} {\bibfnamefont {M.~J.}\ \bibnamefont
  {Dolan}}, \bibinfo {author} {\bibfnamefont {S.~A.}\ \bibnamefont {Malik}},\
  and\ \bibinfo {author} {\bibfnamefont {C.}~\bibnamefont {McCabe}},\ }\href
  {https://doi.org/10.1007/JHEP01(2015)037} {\bibfield  {journal} {\bibinfo
  {journal} {JHEP}\ }\textbf {\bibinfo {volume} {01}},\ \bibinfo {pages}
  {037}},\ \Eprint {https://arxiv.org/abs/1407.8257} {arXiv:1407.8257 [hep-ph]}
  \BibitemShut {NoStop}%
\bibitem [{\citenamefont {D'Eramo}\ \emph {et~al.}(2016)\citenamefont
  {D'Eramo}, \citenamefont {Kavanagh},\ and\ \citenamefont
  {Panci}}]{DEramo:2016gos}%
  \BibitemOpen
  \bibfield  {author} {\bibinfo {author} {\bibfnamefont {F.}~\bibnamefont
  {D'Eramo}}, \bibinfo {author} {\bibfnamefont {B.~J.}\ \bibnamefont
  {Kavanagh}},\ and\ \bibinfo {author} {\bibfnamefont {P.}~\bibnamefont
  {Panci}},\ }\href {https://doi.org/10.1007/JHEP08(2016)111} {\bibfield
  {journal} {\bibinfo  {journal} {JHEP}\ }\textbf {\bibinfo {volume} {08}},\
  \bibinfo {pages} {111}},\ \Eprint {https://arxiv.org/abs/1605.04917}
  {arXiv:1605.04917 [hep-ph]} \BibitemShut {NoStop}%
\bibitem [{\citenamefont {Duerr}\ \emph {et~al.}(2016)\citenamefont {Duerr},
  \citenamefont {Kahlhoefer}, \citenamefont {Schmidt-Hoberg}, \citenamefont
  {Schwetz},\ and\ \citenamefont {Vogl}}]{Duerr:2016tmh}%
  \BibitemOpen
  \bibfield  {author} {\bibinfo {author} {\bibfnamefont {M.}~\bibnamefont
  {Duerr}}, \bibinfo {author} {\bibfnamefont {F.}~\bibnamefont {Kahlhoefer}},
  \bibinfo {author} {\bibfnamefont {K.}~\bibnamefont {Schmidt-Hoberg}},
  \bibinfo {author} {\bibfnamefont {T.}~\bibnamefont {Schwetz}},\ and\ \bibinfo
  {author} {\bibfnamefont {S.}~\bibnamefont {Vogl}},\ }\href
  {https://doi.org/10.1007/JHEP09(2016)042} {\bibfield  {journal} {\bibinfo
  {journal} {JHEP}\ }\textbf {\bibinfo {volume} {09}},\ \bibinfo {pages}
  {042}},\ \Eprint {https://arxiv.org/abs/1606.07609} {arXiv:1606.07609
  [hep-ph]} \BibitemShut {NoStop}%
\bibitem [{\citenamefont {Cohen}\ \emph {et~al.}(2022)\citenamefont {Cohen},
  \citenamefont {Doss},\ and\ \citenamefont {Lu}}]{Cohen:2021gdw}%
  \BibitemOpen
  \bibfield  {author} {\bibinfo {author} {\bibfnamefont {T.}~\bibnamefont
  {Cohen}}, \bibinfo {author} {\bibfnamefont {J.}~\bibnamefont {Doss}},\ and\
  \bibinfo {author} {\bibfnamefont {X.}~\bibnamefont {Lu}},\ }\href
  {https://doi.org/10.1007/JHEP04(2022)155} {\bibfield  {journal} {\bibinfo
  {journal} {JHEP}\ }\textbf {\bibinfo {volume} {04}},\ \bibinfo {pages}
  {155}},\ \Eprint {https://arxiv.org/abs/2111.09895} {arXiv:2111.09895
  [hep-ph]} \BibitemShut {NoStop}%
\bibitem [{\citenamefont {Goodman}\ \emph {et~al.}(2010)\citenamefont
  {Goodman}, \citenamefont {Ibe}, \citenamefont {Rajaraman}, \citenamefont
  {Shepherd}, \citenamefont {Tait},\ and\ \citenamefont {Yu}}]{Goodman:2010ku}%
  \BibitemOpen
  \bibfield  {author} {\bibinfo {author} {\bibfnamefont {J.}~\bibnamefont
  {Goodman}}, \bibinfo {author} {\bibfnamefont {M.}~\bibnamefont {Ibe}},
  \bibinfo {author} {\bibfnamefont {A.}~\bibnamefont {Rajaraman}}, \bibinfo
  {author} {\bibfnamefont {W.}~\bibnamefont {Shepherd}}, \bibinfo {author}
  {\bibfnamefont {T.~M.~P.}\ \bibnamefont {Tait}},\ and\ \bibinfo {author}
  {\bibfnamefont {H.-B.}\ \bibnamefont {Yu}},\ }\href
  {https://doi.org/10.1103/PhysRevD.82.116010} {\bibfield  {journal} {\bibinfo
  {journal} {Phys. Rev. D}\ }\textbf {\bibinfo {volume} {82}},\ \bibinfo
  {pages} {116010} (\bibinfo {year} {2010})},\ \Eprint
  {https://arxiv.org/abs/1008.1783} {arXiv:1008.1783 [hep-ph]} \BibitemShut
  {NoStop}%
\bibitem [{\citenamefont {Fox}\ \emph {et~al.}(2012)\citenamefont {Fox},
  \citenamefont {Harnik}, \citenamefont {Kopp},\ and\ \citenamefont
  {Tsai}}]{Fox:2011pm}%
  \BibitemOpen
  \bibfield  {author} {\bibinfo {author} {\bibfnamefont {P.~J.}\ \bibnamefont
  {Fox}}, \bibinfo {author} {\bibfnamefont {R.}~\bibnamefont {Harnik}},
  \bibinfo {author} {\bibfnamefont {J.}~\bibnamefont {Kopp}},\ and\ \bibinfo
  {author} {\bibfnamefont {Y.}~\bibnamefont {Tsai}},\ }\href
  {https://doi.org/10.1103/PhysRevD.85.056011} {\bibfield  {journal} {\bibinfo
  {journal} {Phys. Rev. D}\ }\textbf {\bibinfo {volume} {85}},\ \bibinfo
  {pages} {056011} (\bibinfo {year} {2012})},\ \Eprint
  {https://arxiv.org/abs/1109.4398} {arXiv:1109.4398 [hep-ph]} \BibitemShut
  {NoStop}%
\bibitem [{\citenamefont {Lin}\ \emph {et~al.}(2013)\citenamefont {Lin},
  \citenamefont {Kolb},\ and\ \citenamefont {Wang}}]{Lin:2013sca}%
  \BibitemOpen
  \bibfield  {author} {\bibinfo {author} {\bibfnamefont {T.}~\bibnamefont
  {Lin}}, \bibinfo {author} {\bibfnamefont {E.~W.}\ \bibnamefont {Kolb}},\ and\
  \bibinfo {author} {\bibfnamefont {L.-T.}\ \bibnamefont {Wang}},\ }\href
  {https://doi.org/10.1103/PhysRevD.88.063510} {\bibfield  {journal} {\bibinfo
  {journal} {Phys. Rev. D}\ }\textbf {\bibinfo {volume} {88}},\ \bibinfo
  {pages} {063510} (\bibinfo {year} {2013})},\ \Eprint
  {https://arxiv.org/abs/1303.6638} {arXiv:1303.6638 [hep-ph]} \BibitemShut
  {NoStop}%
\bibitem [{\citenamefont {Haisch}\ and\ \citenamefont
  {Re}(2015)}]{Haisch:2015ioa}%
  \BibitemOpen
  \bibfield  {author} {\bibinfo {author} {\bibfnamefont {U.}~\bibnamefont
  {Haisch}}\ and\ \bibinfo {author} {\bibfnamefont {E.}~\bibnamefont {Re}},\
  }\href {https://doi.org/10.1007/JHEP06(2015)078} {\bibfield  {journal}
  {\bibinfo  {journal} {JHEP}\ }\textbf {\bibinfo {volume} {06}},\ \bibinfo
  {pages} {078}},\ \Eprint {https://arxiv.org/abs/1503.00691} {arXiv:1503.00691
  [hep-ph]} \BibitemShut {NoStop}%
\bibitem [{\citenamefont {Bell}\ \emph {et~al.}(2016)\citenamefont {Bell},
  \citenamefont {Cai},\ and\ \citenamefont {Leane}}]{Bell:2015rdw}%
  \BibitemOpen
  \bibfield  {author} {\bibinfo {author} {\bibfnamefont {N.~F.}\ \bibnamefont
  {Bell}}, \bibinfo {author} {\bibfnamefont {Y.}~\bibnamefont {Cai}},\ and\
  \bibinfo {author} {\bibfnamefont {R.~K.}\ \bibnamefont {Leane}},\ }\href
  {https://doi.org/10.1088/1475-7516/2016/01/051} {\bibfield  {journal}
  {\bibinfo  {journal} {JCAP}\ }\textbf {\bibinfo {volume} {01}},\ \bibinfo
  {pages} {051}},\ \Eprint {https://arxiv.org/abs/1512.00476} {arXiv:1512.00476
  [hep-ph]} \BibitemShut {NoStop}%
\bibitem [{\citenamefont {Bai}\ and\ \citenamefont {Tait}(2013)}]{Bai:2012xg}%
  \BibitemOpen
  \bibfield  {author} {\bibinfo {author} {\bibfnamefont {Y.}~\bibnamefont
  {Bai}}\ and\ \bibinfo {author} {\bibfnamefont {T.~M.~P.}\ \bibnamefont
  {Tait}},\ }\href {https://doi.org/10.1016/j.physletb.2013.05.057} {\bibfield
  {journal} {\bibinfo  {journal} {Phys. Lett. B}\ }\textbf {\bibinfo {volume}
  {723}},\ \bibinfo {pages} {384} (\bibinfo {year} {2013})},\ \Eprint
  {https://arxiv.org/abs/1208.4361} {arXiv:1208.4361 [hep-ph]} \BibitemShut
  {NoStop}%
\bibitem [{\citenamefont {Bell}\ \emph {et~al.}(2012)\citenamefont {Bell},
  \citenamefont {Dent}, \citenamefont {Galea}, \citenamefont {Jacques},
  \citenamefont {Krauss},\ and\ \citenamefont {Weiler}}]{Bell:2012rg}%
  \BibitemOpen
  \bibfield  {author} {\bibinfo {author} {\bibfnamefont {N.~F.}\ \bibnamefont
  {Bell}}, \bibinfo {author} {\bibfnamefont {J.~B.}\ \bibnamefont {Dent}},
  \bibinfo {author} {\bibfnamefont {A.~J.}\ \bibnamefont {Galea}}, \bibinfo
  {author} {\bibfnamefont {T.~D.}\ \bibnamefont {Jacques}}, \bibinfo {author}
  {\bibfnamefont {L.~M.}\ \bibnamefont {Krauss}},\ and\ \bibinfo {author}
  {\bibfnamefont {T.~J.}\ \bibnamefont {Weiler}},\ }\href
  {https://doi.org/10.1103/PhysRevD.86.096011} {\bibfield  {journal} {\bibinfo
  {journal} {Phys. Rev. D}\ }\textbf {\bibinfo {volume} {86}},\ \bibinfo
  {pages} {096011} (\bibinfo {year} {2012})},\ \Eprint
  {https://arxiv.org/abs/1209.0231} {arXiv:1209.0231 [hep-ph]} \BibitemShut
  {NoStop}%
\bibitem [{\citenamefont {Carpenter}\ \emph {et~al.}(2013)\citenamefont
  {Carpenter}, \citenamefont {Nelson}, \citenamefont {Shimmin}, \citenamefont
  {Tait},\ and\ \citenamefont {Whiteson}}]{Carpenter:2012rg}%
  \BibitemOpen
  \bibfield  {author} {\bibinfo {author} {\bibfnamefont {L.~M.}\ \bibnamefont
  {Carpenter}}, \bibinfo {author} {\bibfnamefont {A.}~\bibnamefont {Nelson}},
  \bibinfo {author} {\bibfnamefont {C.}~\bibnamefont {Shimmin}}, \bibinfo
  {author} {\bibfnamefont {T.~M.~P.}\ \bibnamefont {Tait}},\ and\ \bibinfo
  {author} {\bibfnamefont {D.}~\bibnamefont {Whiteson}},\ }\href
  {https://doi.org/10.1103/PhysRevD.87.074005} {\bibfield  {journal} {\bibinfo
  {journal} {Phys. Rev. D}\ }\textbf {\bibinfo {volume} {87}},\ \bibinfo
  {pages} {074005} (\bibinfo {year} {2013})},\ \Eprint
  {https://arxiv.org/abs/1212.3352} {arXiv:1212.3352 [hep-ex]} \BibitemShut
  {NoStop}%
\bibitem [{\citenamefont {Petrov}\ and\ \citenamefont
  {Shepherd}(2014)}]{Petrov:2013nia}%
  \BibitemOpen
  \bibfield  {author} {\bibinfo {author} {\bibfnamefont {A.~A.}\ \bibnamefont
  {Petrov}}\ and\ \bibinfo {author} {\bibfnamefont {W.}~\bibnamefont
  {Shepherd}},\ }\href {https://doi.org/10.1016/j.physletb.2014.01.051}
  {\bibfield  {journal} {\bibinfo  {journal} {Phys. Lett. B}\ }\textbf
  {\bibinfo {volume} {730}},\ \bibinfo {pages} {178} (\bibinfo {year}
  {2014})},\ \Eprint {https://arxiv.org/abs/1311.1511} {arXiv:1311.1511
  [hep-ph]} \BibitemShut {NoStop}%
\bibitem [{\citenamefont {Berlin}\ \emph {et~al.}(2014)\citenamefont {Berlin},
  \citenamefont {Lin},\ and\ \citenamefont {Wang}}]{Berlin:2014cfa}%
  \BibitemOpen
  \bibfield  {author} {\bibinfo {author} {\bibfnamefont {A.}~\bibnamefont
  {Berlin}}, \bibinfo {author} {\bibfnamefont {T.}~\bibnamefont {Lin}},\ and\
  \bibinfo {author} {\bibfnamefont {L.-T.}\ \bibnamefont {Wang}},\ }\href
  {https://doi.org/10.1007/JHEP06(2014)078} {\bibfield  {journal} {\bibinfo
  {journal} {JHEP}\ }\textbf {\bibinfo {volume} {06}},\ \bibinfo {pages}
  {078}},\ \Eprint {https://arxiv.org/abs/1402.7074} {arXiv:1402.7074 [hep-ph]}
  \BibitemShut {NoStop}%
\bibitem [{\citenamefont {Aad}\ \emph {et~al.}(2021)\citenamefont {Aad} \emph
  {et~al.}}]{ATLAS:2021kxv}%
  \BibitemOpen
  \bibfield  {author} {\bibinfo {author} {\bibfnamefont {G.}~\bibnamefont
  {Aad}} \emph {et~al.} (\bibinfo {collaboration} {ATLAS}),\ }\href
  {https://doi.org/10.1103/PhysRevD.103.112006} {\bibfield  {journal} {\bibinfo
   {journal} {Phys. Rev. D}\ }\textbf {\bibinfo {volume} {103}},\ \bibinfo
  {pages} {112006} (\bibinfo {year} {2021})},\ \Eprint
  {https://arxiv.org/abs/2102.10874} {arXiv:2102.10874 [hep-ex]} \BibitemShut
  {NoStop}%
\bibitem [{\citenamefont {Strassler}\ and\ \citenamefont
  {Zurek}(2007)}]{Strassler:2006im}%
  \BibitemOpen
  \bibfield  {author} {\bibinfo {author} {\bibfnamefont {M.~J.}\ \bibnamefont
  {Strassler}}\ and\ \bibinfo {author} {\bibfnamefont {K.~M.}\ \bibnamefont
  {Zurek}},\ }\href {https://doi.org/10.1016/j.physletb.2007.06.055} {\bibfield
   {journal} {\bibinfo  {journal} {Phys. Lett. B}\ }\textbf {\bibinfo {volume}
  {651}},\ \bibinfo {pages} {374} (\bibinfo {year} {2007})},\ \Eprint
  {https://arxiv.org/abs/hep-ph/0604261} {arXiv:hep-ph/0604261} \BibitemShut
  {NoStop}%
\bibitem [{\citenamefont {Han}\ \emph {et~al.}(2008)\citenamefont {Han},
  \citenamefont {Si}, \citenamefont {Zurek},\ and\ \citenamefont
  {Strassler}}]{Han:2007ae}%
  \BibitemOpen
  \bibfield  {author} {\bibinfo {author} {\bibfnamefont {T.}~\bibnamefont
  {Han}}, \bibinfo {author} {\bibfnamefont {Z.}~\bibnamefont {Si}}, \bibinfo
  {author} {\bibfnamefont {K.~M.}\ \bibnamefont {Zurek}},\ and\ \bibinfo
  {author} {\bibfnamefont {M.~J.}\ \bibnamefont {Strassler}},\ }\href
  {https://doi.org/10.1088/1126-6708/2008/07/008} {\bibfield  {journal}
  {\bibinfo  {journal} {JHEP}\ }\textbf {\bibinfo {volume} {07}},\ \bibinfo
  {pages} {008}},\ \Eprint {https://arxiv.org/abs/0712.2041} {arXiv:0712.2041
  [hep-ph]} \BibitemShut {NoStop}%
\bibitem [{\citenamefont {Holdom}(1986)}]{Holdom:1985ag}%
  \BibitemOpen
  \bibfield  {author} {\bibinfo {author} {\bibfnamefont {B.}~\bibnamefont
  {Holdom}},\ }\href {https://doi.org/10.1016/0370-2693(86)91377-8} {\bibfield
  {journal} {\bibinfo  {journal} {Phys. Lett. B}\ }\textbf {\bibinfo {volume}
  {166}},\ \bibinfo {pages} {196} (\bibinfo {year} {1986})}\BibitemShut
  {NoStop}%
\bibitem [{\citenamefont {Gupta}\ \emph {et~al.}(2015)\citenamefont {Gupta},
  \citenamefont {Primulando},\ and\ \citenamefont {Saraswat}}]{Gupta:2015lfa}%
  \BibitemOpen
  \bibfield  {author} {\bibinfo {author} {\bibfnamefont {A.}~\bibnamefont
  {Gupta}}, \bibinfo {author} {\bibfnamefont {R.}~\bibnamefont {Primulando}},\
  and\ \bibinfo {author} {\bibfnamefont {P.}~\bibnamefont {Saraswat}},\ }\href
  {https://doi.org/10.1007/JHEP09(2015)079} {\bibfield  {journal} {\bibinfo
  {journal} {JHEP}\ }\textbf {\bibinfo {volume} {09}},\ \bibinfo {pages}
  {079}},\ \Eprint {https://arxiv.org/abs/1504.01385} {arXiv:1504.01385
  [hep-ph]} \BibitemShut {NoStop}%
\bibitem [{\citenamefont {Bai}\ \emph {et~al.}(2015)\citenamefont {Bai},
  \citenamefont {Bourbeau},\ and\ \citenamefont {Lin}}]{Bai:2015nfa}%
  \BibitemOpen
  \bibfield  {author} {\bibinfo {author} {\bibfnamefont {Y.}~\bibnamefont
  {Bai}}, \bibinfo {author} {\bibfnamefont {J.}~\bibnamefont {Bourbeau}},\ and\
  \bibinfo {author} {\bibfnamefont {T.}~\bibnamefont {Lin}},\ }\href
  {https://doi.org/10.1007/JHEP06(2015)205} {\bibfield  {journal} {\bibinfo
  {journal} {JHEP}\ }\textbf {\bibinfo {volume} {06}},\ \bibinfo {pages}
  {205}},\ \Eprint {https://arxiv.org/abs/1504.01395} {arXiv:1504.01395
  [hep-ph]} \BibitemShut {NoStop}%
\bibitem [{\citenamefont {Tsai}\ \emph {et~al.}(2016)\citenamefont {Tsai},
  \citenamefont {Wang},\ and\ \citenamefont {Zhao}}]{Tsai:2015ugz}%
  \BibitemOpen
  \bibfield  {author} {\bibinfo {author} {\bibfnamefont {Y.}~\bibnamefont
  {Tsai}}, \bibinfo {author} {\bibfnamefont {L.-T.}\ \bibnamefont {Wang}},\
  and\ \bibinfo {author} {\bibfnamefont {Y.}~\bibnamefont {Zhao}},\ }\href
  {https://doi.org/10.1103/PhysRevD.93.035024} {\bibfield  {journal} {\bibinfo
  {journal} {Phys. Rev. D}\ }\textbf {\bibinfo {volume} {93}},\ \bibinfo
  {pages} {035024} (\bibinfo {year} {2016})},\ \Eprint
  {https://arxiv.org/abs/1511.07433} {arXiv:1511.07433 [hep-ph]} \BibitemShut
  {NoStop}%
\bibitem [{\citenamefont {Lindner}\ \emph {et~al.}(2016)\citenamefont
  {Lindner}, \citenamefont {Queiroz},\ and\ \citenamefont
  {Rodejohann}}]{Lindner:2016lpp}%
  \BibitemOpen
  \bibfield  {author} {\bibinfo {author} {\bibfnamefont {M.}~\bibnamefont
  {Lindner}}, \bibinfo {author} {\bibfnamefont {F.~S.}\ \bibnamefont
  {Queiroz}},\ and\ \bibinfo {author} {\bibfnamefont {W.}~\bibnamefont
  {Rodejohann}},\ }\href {https://doi.org/10.1016/j.physletb.2016.08.068}
  {\bibfield  {journal} {\bibinfo  {journal} {Phys. Lett. B}\ }\textbf
  {\bibinfo {volume} {762}},\ \bibinfo {pages} {190} (\bibinfo {year}
  {2016})},\ \Eprint {https://arxiv.org/abs/1604.07419} {arXiv:1604.07419
  [hep-ph]} \BibitemShut {NoStop}%
\bibitem [{\citenamefont {Buschmann}\ \emph {et~al.}(2016)\citenamefont
  {Buschmann}, \citenamefont {El~Hedri}, \citenamefont {Kaminska},
  \citenamefont {Liu}, \citenamefont {de~Vries}, \citenamefont {Wang},
  \citenamefont {Yu},\ and\ \citenamefont {Zurita}}]{Buschmann:2016hkc}%
  \BibitemOpen
  \bibfield  {author} {\bibinfo {author} {\bibfnamefont {M.}~\bibnamefont
  {Buschmann}}, \bibinfo {author} {\bibfnamefont {S.}~\bibnamefont {El~Hedri}},
  \bibinfo {author} {\bibfnamefont {A.}~\bibnamefont {Kaminska}}, \bibinfo
  {author} {\bibfnamefont {J.}~\bibnamefont {Liu}}, \bibinfo {author}
  {\bibfnamefont {M.}~\bibnamefont {de~Vries}}, \bibinfo {author}
  {\bibfnamefont {X.-P.}\ \bibnamefont {Wang}}, \bibinfo {author}
  {\bibfnamefont {F.}~\bibnamefont {Yu}},\ and\ \bibinfo {author}
  {\bibfnamefont {J.}~\bibnamefont {Zurita}},\ }\href
  {https://doi.org/10.1007/JHEP09(2016)033} {\bibfield  {journal} {\bibinfo
  {journal} {JHEP}\ }\textbf {\bibinfo {volume} {09}},\ \bibinfo {pages}
  {033}},\ \Eprint {https://arxiv.org/abs/1605.08056} {arXiv:1605.08056
  [hep-ph]} \BibitemShut {NoStop}%
\bibitem [{\citenamefont {Kim}\ \emph {et~al.}(2018)\citenamefont {Kim},
  \citenamefont {Lee}, \citenamefont {Park},\ and\ \citenamefont
  {Zhang}}]{Kim:2016fdv}%
  \BibitemOpen
  \bibfield  {author} {\bibinfo {author} {\bibfnamefont {M.}~\bibnamefont
  {Kim}}, \bibinfo {author} {\bibfnamefont {H.-S.}\ \bibnamefont {Lee}},
  \bibinfo {author} {\bibfnamefont {M.}~\bibnamefont {Park}},\ and\ \bibinfo
  {author} {\bibfnamefont {M.}~\bibnamefont {Zhang}},\ }\href
  {https://doi.org/10.1103/PhysRevD.98.055027} {\bibfield  {journal} {\bibinfo
  {journal} {Phys. Rev. D}\ }\textbf {\bibinfo {volume} {98}},\ \bibinfo
  {pages} {055027} (\bibinfo {year} {2018})},\ \Eprint
  {https://arxiv.org/abs/1612.02850} {arXiv:1612.02850 [hep-ph]} \BibitemShut
  {NoStop}%
\bibitem [{\citenamefont {Nam}(2022)}]{Nam:2021bsf}%
  \BibitemOpen
  \bibfield  {author} {\bibinfo {author} {\bibfnamefont {C.~H.}\ \bibnamefont
  {Nam}},\ }\href {https://doi.org/10.1103/PhysRevD.105.075015} {\bibfield
  {journal} {\bibinfo  {journal} {Phys. Rev. D}\ }\textbf {\bibinfo {volume}
  {105}},\ \bibinfo {pages} {075015} (\bibinfo {year} {2022})},\ \Eprint
  {https://arxiv.org/abs/2112.10446} {arXiv:2112.10446 [hep-ph]} \BibitemShut
  {NoStop}%
\bibitem [{\citenamefont {Kim}\ \emph {et~al.}(2019)\citenamefont {Kim},
  \citenamefont {Park},\ and\ \citenamefont {Shin}}]{Kim:2019had}%
  \BibitemOpen
  \bibfield  {author} {\bibinfo {author} {\bibfnamefont {D.}~\bibnamefont
  {Kim}}, \bibinfo {author} {\bibfnamefont {J.-C.}\ \bibnamefont {Park}},\ and\
  \bibinfo {author} {\bibfnamefont {S.}~\bibnamefont {Shin}},\ }\href
  {https://doi.org/10.1103/PhysRevD.100.035033} {\bibfield  {journal} {\bibinfo
   {journal} {Phys. Rev. D}\ }\textbf {\bibinfo {volume} {100}},\ \bibinfo
  {pages} {035033} (\bibinfo {year} {2019})},\ \Eprint
  {https://arxiv.org/abs/1903.05087} {arXiv:1903.05087 [hep-ph]} \BibitemShut
  {NoStop}%
\bibitem [{\citenamefont {Bell}\ \emph
  {et~al.}(2017{\natexlab{a}})\citenamefont {Bell}, \citenamefont {Cai},
  \citenamefont {Dent}, \citenamefont {Leane},\ and\ \citenamefont
  {Weiler}}]{Bell:2017irk}%
  \BibitemOpen
  \bibfield  {author} {\bibinfo {author} {\bibfnamefont {N.~F.}\ \bibnamefont
  {Bell}}, \bibinfo {author} {\bibfnamefont {Y.}~\bibnamefont {Cai}}, \bibinfo
  {author} {\bibfnamefont {J.~B.}\ \bibnamefont {Dent}}, \bibinfo {author}
  {\bibfnamefont {R.~K.}\ \bibnamefont {Leane}},\ and\ \bibinfo {author}
  {\bibfnamefont {T.~J.}\ \bibnamefont {Weiler}},\ }\href
  {https://doi.org/10.1103/PhysRevD.96.023011} {\bibfield  {journal} {\bibinfo
  {journal} {Phys. Rev. D}\ }\textbf {\bibinfo {volume} {96}},\ \bibinfo
  {pages} {023011} (\bibinfo {year} {2017}{\natexlab{a}})},\ \Eprint
  {https://arxiv.org/abs/1705.01105} {arXiv:1705.01105 [hep-ph]} \BibitemShut
  {NoStop}%
\bibitem [{ATL(2023)}]{ATLAS:2023tmv}%
  \BibitemOpen
  \href {https://cds.cern.ch/record/2870113} {\emph {\bibinfo {title} {{Search
  for a new leptonically decaying neutral vector boson in association with
  missing transverse energy in proton–proton collisions at $\sqrt{s}=13~$TeV
  with the ATLAS detector}}}},\ \bibinfo {type} {Tech. Rep.}\ (\bibinfo
  {institution} {CERN},\ \bibinfo {address} {Geneva},\ \bibinfo {year} {2023})\
  \bibinfo {note} {all figures including auxiliary figures are available at
  https://atlas.web.cern.ch/Atlas/GROUPS/PHYSICS/CONFNOTES/ATLAS-CONF-2023-045}\BibitemShut
  {NoStop}%
\bibitem [{\citenamefont {Elgammal}\ \emph {et~al.}(2023)\citenamefont
  {Elgammal}, \citenamefont {Louka}, \citenamefont {Ellithi},\ and\
  \citenamefont {Hussein}}]{Elgammal:2021rne}%
  \BibitemOpen
  \bibfield  {author} {\bibinfo {author} {\bibfnamefont {S.}~\bibnamefont
  {Elgammal}}, \bibinfo {author} {\bibfnamefont {M.}~\bibnamefont {Louka}},
  \bibinfo {author} {\bibfnamefont {A.~Y.}\ \bibnamefont {Ellithi}},\ and\
  \bibinfo {author} {\bibfnamefont {M.~T.}\ \bibnamefont {Hussein}},\ }\href
  {https://doi.org/10.1140/epjp/s13360-023-04088-w} {\bibfield  {journal}
  {\bibinfo  {journal} {Eur. Phys. J. Plus}\ }\textbf {\bibinfo {volume}
  {138}},\ \bibinfo {pages} {548} (\bibinfo {year} {2023})},\ \Eprint
  {https://arxiv.org/abs/2109.11274} {arXiv:2109.11274 [hep-ex]} \BibitemShut
  {NoStop}%
\bibitem [{\citenamefont {Aad}\ \emph {et~al.}(2020)\citenamefont {Aad} \emph
  {et~al.}}]{ATLAS:2019lff}%
  \BibitemOpen
  \bibfield  {author} {\bibinfo {author} {\bibfnamefont {G.}~\bibnamefont
  {Aad}} \emph {et~al.} (\bibinfo {collaboration} {ATLAS}),\ }\href
  {https://doi.org/10.1140/epjc/s10052-019-7594-6} {\bibfield  {journal}
  {\bibinfo  {journal} {Eur. Phys. J. C}\ }\textbf {\bibinfo {volume} {80}},\
  \bibinfo {pages} {123} (\bibinfo {year} {2020})},\ \Eprint
  {https://arxiv.org/abs/1908.08215} {arXiv:1908.08215 [hep-ex]} \BibitemShut
  {NoStop}%
\bibitem [{\citenamefont {Sirunyan}\ \emph
  {et~al.}(2021{\natexlab{a}})\citenamefont {Sirunyan} \emph
  {et~al.}}]{CMS:2020bfa}%
  \BibitemOpen
  \bibfield  {author} {\bibinfo {author} {\bibfnamefont {A.~M.}\ \bibnamefont
  {Sirunyan}} \emph {et~al.} (\bibinfo {collaboration} {CMS}),\ }\href
  {https://doi.org/10.1007/JHEP04(2021)123} {\bibfield  {journal} {\bibinfo
  {journal} {JHEP}\ }\textbf {\bibinfo {volume} {04}},\ \bibinfo {pages}
  {123}},\ \Eprint {https://arxiv.org/abs/2012.08600} {arXiv:2012.08600
  [hep-ex]} \BibitemShut {NoStop}%
\bibitem [{\citenamefont {San}\ \emph {et~al.}(2022)\citenamefont {San},
  \citenamefont {Perelstein},\ and\ \citenamefont {Tanedo}}]{San:2021zpeps}%
  \BibitemOpen
  \bibfield  {author} {\bibinfo {author} {\bibfnamefont {Y.~C.}\ \bibnamefont
  {San}}, \bibinfo {author} {\bibfnamefont {M.}~\bibnamefont {Perelstein}},\
  and\ \bibinfo {author} {\bibfnamefont {P.}~\bibnamefont {Tanedo}},\ }\href
  {https://doi.org/10.1103/PhysRevD.106.015027} {\bibfield  {journal} {\bibinfo
   {journal} {Phys. Rev. D}\ }\textbf {\bibinfo {volume} {106}},\ \bibinfo
  {pages} {015027} (\bibinfo {year} {2022})}\BibitemShut {NoStop}%
\bibitem [{\citenamefont {Sirunyan}\ \emph {et~al.}(2020)\citenamefont
  {Sirunyan} \emph {et~al.}}]{CMS:2020epss}%
  \BibitemOpen
  \bibfield  {author} {\bibinfo {author} {\bibfnamefont {A.~M.}\ \bibnamefont
  {Sirunyan}} \emph {et~al.} (\bibinfo {collaboration} {CMS}),\ }\bibfield
  {journal} {\bibinfo  {journal} {Physical Review Letters}\ }\textbf {\bibinfo
  {volume} {124}},\ \href {https://doi.org/10.1103/physrevlett.124.131802}
  {10.1103/physrevlett.124.131802} (\bibinfo {year} {2020}),\ \Eprint
  {https://arxiv.org/abs/1912.04776} {arXiv:1912.04776 [hep-ex]} \BibitemShut
  {NoStop}%
\bibitem [{\citenamefont {Curtin}\ \emph {et~al.}(2015)\citenamefont {Curtin},
  \citenamefont {Essig}, \citenamefont {Gori},\ and\ \citenamefont
  {Shelton}}]{Curtin:2014cca}%
  \BibitemOpen
  \bibfield  {author} {\bibinfo {author} {\bibfnamefont {D.}~\bibnamefont
  {Curtin}}, \bibinfo {author} {\bibfnamefont {R.}~\bibnamefont {Essig}},
  \bibinfo {author} {\bibfnamefont {S.}~\bibnamefont {Gori}},\ and\ \bibinfo
  {author} {\bibfnamefont {J.}~\bibnamefont {Shelton}},\ }\href
  {https://doi.org/10.1007/JHEP02(2015)157} {\bibfield  {journal} {\bibinfo
  {journal} {JHEP}\ }\textbf {\bibinfo {volume} {02}},\ \bibinfo {pages}
  {157}},\ \Eprint {https://arxiv.org/abs/1412.0018} {arXiv:1412.0018 [hep-ph]}
  \BibitemShut {NoStop}%
\bibitem [{\citenamefont {Bell}\ \emph
  {et~al.}(2017{\natexlab{b}})\citenamefont {Bell}, \citenamefont {Cai},\ and\
  \citenamefont {Leane}}]{Bell:2016uhg}%
  \BibitemOpen
  \bibfield  {author} {\bibinfo {author} {\bibfnamefont {N.~F.}\ \bibnamefont
  {Bell}}, \bibinfo {author} {\bibfnamefont {Y.}~\bibnamefont {Cai}},\ and\
  \bibinfo {author} {\bibfnamefont {R.~K.}\ \bibnamefont {Leane}},\ }\href
  {https://doi.org/10.1088/1475-7516/2017/01/039} {\bibfield  {journal}
  {\bibinfo  {journal} {JCAP}\ }\textbf {\bibinfo {volume} {01}},\ \bibinfo
  {pages} {039}},\ \Eprint {https://arxiv.org/abs/1610.03063} {arXiv:1610.03063
  [hep-ph]} \BibitemShut {NoStop}%
\bibitem [{\citenamefont {Aad}\ \emph {et~al.}(2019)\citenamefont {Aad} \emph
  {et~al.}}]{ATLAS:2019erb}%
  \BibitemOpen
  \bibfield  {author} {\bibinfo {author} {\bibfnamefont {G.}~\bibnamefont
  {Aad}} \emph {et~al.} (\bibinfo {collaboration} {ATLAS}),\ }\href
  {https://doi.org/10.1016/j.physletb.2019.07.016} {\bibfield  {journal}
  {\bibinfo  {journal} {Phys. Lett. B}\ }\textbf {\bibinfo {volume} {796}},\
  \bibinfo {pages} {68} (\bibinfo {year} {2019})},\ \Eprint
  {https://arxiv.org/abs/1903.06248} {arXiv:1903.06248 [hep-ex]} \BibitemShut
  {NoStop}%
\bibitem [{\citenamefont {Sirunyan}\ \emph
  {et~al.}(2021{\natexlab{b}})\citenamefont {Sirunyan} \emph
  {et~al.}}]{CMS:2021ctt}%
  \BibitemOpen
  \bibfield  {author} {\bibinfo {author} {\bibfnamefont {A.~M.}\ \bibnamefont
  {Sirunyan}} \emph {et~al.} (\bibinfo {collaboration} {CMS}),\ }\href
  {https://doi.org/10.1007/JHEP07(2021)208} {\bibfield  {journal} {\bibinfo
  {journal} {JHEP}\ }\textbf {\bibinfo {volume} {07}},\ \bibinfo {pages}
  {208}},\ \Eprint {https://arxiv.org/abs/2103.02708} {arXiv:2103.02708
  [hep-ex]} \BibitemShut {NoStop}%
\bibitem [{\citenamefont {Buschmann}\ \emph {et~al.}(2015)\citenamefont
  {Buschmann}, \citenamefont {Kopp}, \citenamefont {Liu},\ and\ \citenamefont
  {Machado}}]{Buschmann:2015awa}%
  \BibitemOpen
  \bibfield  {author} {\bibinfo {author} {\bibfnamefont {M.}~\bibnamefont
  {Buschmann}}, \bibinfo {author} {\bibfnamefont {J.}~\bibnamefont {Kopp}},
  \bibinfo {author} {\bibfnamefont {J.}~\bibnamefont {Liu}},\ and\ \bibinfo
  {author} {\bibfnamefont {P.~A.~N.}\ \bibnamefont {Machado}},\ }\href
  {https://doi.org/10.1007/JHEP07(2015)045} {\bibfield  {journal} {\bibinfo
  {journal} {JHEP}\ }\textbf {\bibinfo {volume} {07}},\ \bibinfo {pages}
  {045}},\ \Eprint {https://arxiv.org/abs/1505.07459} {arXiv:1505.07459
  [hep-ph]} \BibitemShut {NoStop}%
\bibitem [{\citenamefont {de~Favereau}\ \emph {et~al.}(2014)\citenamefont
  {de~Favereau}, \citenamefont {Delaere}, \citenamefont {Demin}, \citenamefont
  {Giammanco}, \citenamefont {Lema\^\i{}tre}, \citenamefont {Mertens},\ and\
  \citenamefont {Selvaggi}}]{deFavereau:2013fsa}%
  \BibitemOpen
  \bibfield  {author} {\bibinfo {author} {\bibfnamefont {J.}~\bibnamefont
  {de~Favereau}}, \bibinfo {author} {\bibfnamefont {C.}~\bibnamefont
  {Delaere}}, \bibinfo {author} {\bibfnamefont {P.}~\bibnamefont {Demin}},
  \bibinfo {author} {\bibfnamefont {A.}~\bibnamefont {Giammanco}}, \bibinfo
  {author} {\bibfnamefont {V.}~\bibnamefont {Lema\^\i{}tre}}, \bibinfo {author}
  {\bibfnamefont {A.}~\bibnamefont {Mertens}},\ and\ \bibinfo {author}
  {\bibfnamefont {M.}~\bibnamefont {Selvaggi}} (\bibinfo {collaboration}
  {DELPHES 3}),\ }\href {https://doi.org/10.1007/JHEP02(2014)057} {\bibfield
  {journal} {\bibinfo  {journal} {JHEP}\ }\textbf {\bibinfo {volume} {02}},\
  \bibinfo {pages} {057}},\ \Eprint {https://arxiv.org/abs/1307.6346}
  {arXiv:1307.6346 [hep-ex]} \BibitemShut {NoStop}%
\bibitem [{\citenamefont {Alwall}\ \emph {et~al.}(2014)\citenamefont {Alwall},
  \citenamefont {Frederix}, \citenamefont {Frixione}, \citenamefont {Hirschi},
  \citenamefont {Maltoni}, \citenamefont {Mattelaer}, \citenamefont {Shao},
  \citenamefont {Stelzer}, \citenamefont {Torrielli},\ and\ \citenamefont
  {Zaro}}]{Alwall:2014hca}%
  \BibitemOpen
  \bibfield  {author} {\bibinfo {author} {\bibfnamefont {J.}~\bibnamefont
  {Alwall}}, \bibinfo {author} {\bibfnamefont {R.}~\bibnamefont {Frederix}},
  \bibinfo {author} {\bibfnamefont {S.}~\bibnamefont {Frixione}}, \bibinfo
  {author} {\bibfnamefont {V.}~\bibnamefont {Hirschi}}, \bibinfo {author}
  {\bibfnamefont {F.}~\bibnamefont {Maltoni}}, \bibinfo {author} {\bibfnamefont
  {O.}~\bibnamefont {Mattelaer}}, \bibinfo {author} {\bibfnamefont {H.~S.}\
  \bibnamefont {Shao}}, \bibinfo {author} {\bibfnamefont {T.}~\bibnamefont
  {Stelzer}}, \bibinfo {author} {\bibfnamefont {P.}~\bibnamefont {Torrielli}},\
  and\ \bibinfo {author} {\bibfnamefont {M.}~\bibnamefont {Zaro}},\ }\href
  {https://doi.org/10.1007/JHEP07(2014)079} {\bibfield  {journal} {\bibinfo
  {journal} {JHEP}\ }\textbf {\bibinfo {volume} {07}},\ \bibinfo {pages}
  {079}},\ \Eprint {https://arxiv.org/abs/1405.0301} {arXiv:1405.0301 [hep-ph]}
  \BibitemShut {NoStop}%
\bibitem [{\citenamefont {Alloul}\ \emph {et~al.}(2014)\citenamefont {Alloul},
  \citenamefont {Christensen}, \citenamefont {Degrande}, \citenamefont {Duhr},\
  and\ \citenamefont {Fuks}}]{Alloul:2013bka}%
  \BibitemOpen
  \bibfield  {author} {\bibinfo {author} {\bibfnamefont {A.}~\bibnamefont
  {Alloul}}, \bibinfo {author} {\bibfnamefont {N.~D.}\ \bibnamefont
  {Christensen}}, \bibinfo {author} {\bibfnamefont {C.}~\bibnamefont
  {Degrande}}, \bibinfo {author} {\bibfnamefont {C.}~\bibnamefont {Duhr}},\
  and\ \bibinfo {author} {\bibfnamefont {B.}~\bibnamefont {Fuks}},\ }\href
  {https://doi.org/10.1016/j.cpc.2014.04.012} {\bibfield  {journal} {\bibinfo
  {journal} {Comput. Phys. Commun.}\ }\textbf {\bibinfo {volume} {185}},\
  \bibinfo {pages} {2250} (\bibinfo {year} {2014})},\ \Eprint
  {https://arxiv.org/abs/1310.1921} {arXiv:1310.1921 [hep-ph]} \BibitemShut
  {NoStop}%
\bibitem [{\citenamefont {Sj\"ostrand}\ \emph {et~al.}(2015)\citenamefont
  {Sj\"ostrand}, \citenamefont {Ask}, \citenamefont {Christiansen},
  \citenamefont {Corke}, \citenamefont {Desai}, \citenamefont {Ilten},
  \citenamefont {Mrenna}, \citenamefont {Prestel}, \citenamefont {Rasmussen},\
  and\ \citenamefont {Skands}}]{Sjostrand:2014zea}%
  \BibitemOpen
  \bibfield  {author} {\bibinfo {author} {\bibfnamefont {T.}~\bibnamefont
  {Sj\"ostrand}}, \bibinfo {author} {\bibfnamefont {S.}~\bibnamefont {Ask}},
  \bibinfo {author} {\bibfnamefont {J.~R.}\ \bibnamefont {Christiansen}},
  \bibinfo {author} {\bibfnamefont {R.}~\bibnamefont {Corke}}, \bibinfo
  {author} {\bibfnamefont {N.}~\bibnamefont {Desai}}, \bibinfo {author}
  {\bibfnamefont {P.}~\bibnamefont {Ilten}}, \bibinfo {author} {\bibfnamefont
  {S.}~\bibnamefont {Mrenna}}, \bibinfo {author} {\bibfnamefont
  {S.}~\bibnamefont {Prestel}}, \bibinfo {author} {\bibfnamefont {C.~O.}\
  \bibnamefont {Rasmussen}},\ and\ \bibinfo {author} {\bibfnamefont {P.~Z.}\
  \bibnamefont {Skands}},\ }\href {https://doi.org/10.1016/j.cpc.2015.01.024}
  {\bibfield  {journal} {\bibinfo  {journal} {Comput. Phys. Commun.}\ }\textbf
  {\bibinfo {volume} {191}},\ \bibinfo {pages} {159} (\bibinfo {year}
  {2015})},\ \Eprint {https://arxiv.org/abs/1410.3012} {arXiv:1410.3012
  [hep-ph]} \BibitemShut {NoStop}%
\bibitem [{\citenamefont {Cacciari}\ \emph {et~al.}(2012)\citenamefont
  {Cacciari}, \citenamefont {Salam},\ and\ \citenamefont
  {Soyez}}]{Cacciari:2012fj}%
  \BibitemOpen
  \bibfield  {author} {\bibinfo {author} {\bibfnamefont {M.}~\bibnamefont
  {Cacciari}}, \bibinfo {author} {\bibfnamefont {G.~P.}\ \bibnamefont
  {Salam}},\ and\ \bibinfo {author} {\bibfnamefont {G.}~\bibnamefont {Soyez}},\
  }\bibfield  {journal} {\bibinfo  {journal} {The European Physical Journal C}\
  }\textbf {\bibinfo {volume} {72}},\ \href
  {https://doi.org/10.1140/epjc/s10052-012-1896-2}
  {10.1140/epjc/s10052-012-1896-2} (\bibinfo {year} {2012})\BibitemShut
  {NoStop}%
\bibitem [{\citenamefont {Bal\'azs}\ \emph {et~al.}(2017)\citenamefont
  {Bal\'azs} \emph {et~al.}}]{GAMBIT:2017qxg}%
  \BibitemOpen
  \bibfield  {author} {\bibinfo {author} {\bibfnamefont {C.}~\bibnamefont
  {Bal\'azs}} \emph {et~al.} (\bibinfo {collaboration} {GAMBIT}),\ }\href
  {https://doi.org/10.1140/epjc/s10052-017-5285-8} {\bibfield  {journal}
  {\bibinfo  {journal} {Eur. Phys. J. C}\ }\textbf {\bibinfo {volume} {77}},\
  \bibinfo {pages} {795} (\bibinfo {year} {2017})},\ \Eprint
  {https://arxiv.org/abs/1705.07919} {arXiv:1705.07919 [hep-ph]} \BibitemShut
  {NoStop}%
\bibitem [{\citenamefont {Primulando}\ \emph {et~al.}(2020)\citenamefont
  {Primulando}, \citenamefont {Julio},\ and\ \citenamefont
  {Uttayarat}}]{Primulando:2020cbl}%
  \BibitemOpen
  \bibfield  {author} {\bibinfo {author} {\bibfnamefont {R.}~\bibnamefont
  {Primulando}}, \bibinfo {author} {\bibfnamefont {J.}~\bibnamefont {Julio}},\
  and\ \bibinfo {author} {\bibfnamefont {P.}~\bibnamefont {Uttayarat}},\
  }\bibfield  {journal} {\bibinfo  {journal} {Physical Review D}\ }\textbf
  {\bibinfo {volume} {101}},\ \href
  {https://doi.org/10.1103/physrevd.101.055021} {10.1103/physrevd.101.055021}
  (\bibinfo {year} {2020}),\ \Eprint {https://arxiv.org/abs/1912.08533}
  {arXiv:1912.08533 [hep-ph]} \BibitemShut {NoStop}%
\bibitem [{\citenamefont {Tumasyan}\ \emph {et~al.}(2023)\citenamefont
  {Tumasyan} \emph {et~al.}}]{CMS:2022qej}%
  \BibitemOpen
  \bibfield  {author} {\bibinfo {author} {\bibfnamefont {A.}~\bibnamefont
  {Tumasyan}} \emph {et~al.} (\bibinfo {collaboration} {CMS}),\ }\href
  {https://doi.org/10.1007/JHEP05(2023)228} {\bibfield  {journal} {\bibinfo
  {journal} {JHEP}\ }\textbf {\bibinfo {volume} {05}},\ \bibinfo {pages}
  {228}},\ \Eprint {https://arxiv.org/abs/2205.08582} {arXiv:2205.08582
  [hep-ex]} \BibitemShut {NoStop}%
\bibitem [{\citenamefont {Feng}\ \emph {et~al.}(2018)\citenamefont {Feng},
  \citenamefont {Galon}, \citenamefont {Kling},\ and\ \citenamefont
  {Trojanowski}}]{Feng:2017uoz}%
  \BibitemOpen
  \bibfield  {author} {\bibinfo {author} {\bibfnamefont {J.~L.}\ \bibnamefont
  {Feng}}, \bibinfo {author} {\bibfnamefont {I.}~\bibnamefont {Galon}},
  \bibinfo {author} {\bibfnamefont {F.}~\bibnamefont {Kling}},\ and\ \bibinfo
  {author} {\bibfnamefont {S.}~\bibnamefont {Trojanowski}},\ }\href
  {https://doi.org/10.1103/PhysRevD.97.035001} {\bibfield  {journal} {\bibinfo
  {journal} {Phys. Rev. D}\ }\textbf {\bibinfo {volume} {97}},\ \bibinfo
  {pages} {035001} (\bibinfo {year} {2018})},\ \Eprint
  {https://arxiv.org/abs/1708.09389} {arXiv:1708.09389 [hep-ph]} \BibitemShut
  {NoStop}%
\bibitem [{\citenamefont {Curtin}\ \emph {et~al.}(2019)\citenamefont {Curtin}
  \emph {et~al.}}]{Curtin:2018mvb}%
  \BibitemOpen
  \bibfield  {author} {\bibinfo {author} {\bibfnamefont {D.}~\bibnamefont
  {Curtin}} \emph {et~al.},\ }\href {https://doi.org/10.1088/1361-6633/ab28d6}
  {\bibfield  {journal} {\bibinfo  {journal} {Rept. Prog. Phys.}\ }\textbf
  {\bibinfo {volume} {82}},\ \bibinfo {pages} {116201} (\bibinfo {year}
  {2019})},\ \Eprint {https://arxiv.org/abs/1806.07396} {arXiv:1806.07396
  [hep-ph]} \BibitemShut {NoStop}%
\bibitem [{\citenamefont {Sirunyan}\ \emph {et~al.}(2019)\citenamefont
  {Sirunyan} \emph {et~al.}}]{CMS:2019xai}%
  \BibitemOpen
  \bibfield  {author} {\bibinfo {author} {\bibfnamefont {A.~M.}\ \bibnamefont
  {Sirunyan}} \emph {et~al.} (\bibinfo {collaboration} {CMS}),\ }\href
  {https://doi.org/10.1103/PhysRevLett.123.231803} {\bibfield  {journal}
  {\bibinfo  {journal} {Phys. Rev. Lett.}\ }\textbf {\bibinfo {volume} {123}},\
  \bibinfo {pages} {231803} (\bibinfo {year} {2019})},\ \Eprint
  {https://arxiv.org/abs/1905.10331} {arXiv:1905.10331 [hep-ex]} \BibitemShut
  {NoStop}%
\end{thebibliography}%
\bibliographystyle{apsrev4-2}
%%%%%%%%%%%%%%%%%%%%%%%%%%%%%%%%%%%%%%%%%%%%%%%%%%%%%%%%%%%%%%%%%%%%%%%%%%

\end{document}